%% file: CG_Library.tex
\documentclass{amsart}

 \usepackage{amsmath}
\usepackage{amsfonts}
\usepackage{url}
\usepackage{fullpage}
\usepackage{amssymb}
\usepackage[dvips]{graphicx}
\usepackage{color}
\usepackage{float} 
\usepackage{url}
\usepackage{epsfig} 

\newcommand{\transp}{{\scriptscriptstyle \top}}
\newtheorem{thm}{\bf{Theorem}}[section]

\newtheorem{algo}[thm]{\bf{Algorithm}}

\begin{document}

\title[]{{A library to compute the density of the distance between a point and a random
variable uniformly distributed in some sets}}

\maketitle

\begin{center}
Vincent Guigues\\
School of Applied Mathematics, FGV\\
Praia de Botafogo, Rio de Janeiro, Brazil\\ 
{\tt vguigues@fgv.br}
\end{center}

\date{}

\begin{abstract} 
In \cite{guiguesarxivcg2015}, algorithms to compute the density of the
distance to a random variable uniformly distributed in (a) a ball, (b) a disk, (c) a line
segment, or (d) a polygone were introduced. 
For case (d), the algorithm, based on Green's theorem,
has complexity $n$log($n$) where $n$ is the number of 
vertices of the polygone.
In this paper, we present for case (d) another algorithm with the same complexity, based on a triangulation
of the polygone. We also describe an open source library, available at \url{https://github.com/vguigues/Areas_Library},
providing this algorithm as well as the
algorithms from \cite{guiguesarxivcg2015}.\\
\end{abstract}

\par {\textbf{Keywords:} Computational Geometry, Geometric Probability, Distance to a random variable, Uniform distribution, Green's theorem, PSHA.\\

\par MSC2010 subject classifications: 60D05, 65D99, 51N20, 65D30, 86A15.\\

\section{Introduction}

Let $S \subset \mathbb{R}^3$ be a closed and bounded set 
and let $X:\Omega \rightarrow S$ be a 
random variable uniformly distributed in $S$.
Given $P \in \mathbb{R}^3$,
consider random variable $D$
given by the Euclidean distance $D:\Omega \rightarrow \mathbb{R}_{+}$ between $P$ and
$X$, i.e., $D(\omega)=\|\overrightarrow{P X(\omega)}\|_2$ for any $\omega \in \Omega$.

Denoting respectively the density 
and the cumulative distribution function (CDF) 
of $D$ by $f_D(\cdot)$ and $F_D(\cdot)$, we have $f_D(d)=F_D(d)=0$ 
if $d<\displaystyle \min_{Q \in S}\;\|\overrightarrow{PQ}\|_2$ while
$f_D(d)=0$ and $F_D(d)=1$
if $d>\displaystyle \max_{Q \in S}\;\|\overrightarrow{PQ}\|_2$.
For $\displaystyle \min_{Q \in S}\;\|\overrightarrow{PQ}\|_2 \leq d \leq \displaystyle \max_{Q \in S}\;\|\overrightarrow{PQ}\|_2$, we have
$$
F_D(d)=
\mathbb{P}(D\leq d)=\frac{\mu(\mathcal{B}(P,d) \cap S)}{\mu(S)}
$$
where $\mu(A)$ is the Lebesgue measure of the set $A$
and $\mathcal{B}(P, d)$ is the ball of center $P$ and radius $d$.
As a result, the computation of the CDF of $D$ 
amounts to the problem of computing the Lebesgue 
measures  of $S$
and of $\mathcal{B}(P,d) \cap S$ for any $d \in \mathbb{R}_{+}$.

When $S$ is a disk, a ball, or a line segment, 
it is easy to derive analytic expressions for both the CDF
and the density of $D$, see \cite{guiguesarxivcg2015} for details.
When $S$ is a polygone, an algorithm based on
Green's theorem computing exactly the CDF of $D$
is described in \cite{guiguesarxivcg2015}.

The study of these four cases is useful for Probabilistic Seismic Hazard Analysis (PSHA)
to obtain the distribution of the distance between a given location on earth and
the epicenter of an earthquake which, in a given seismic zone, is usually assumed to have a uniform distribution
in that zone modelled as a union of disks, a union of balls, a union of line segments, or the boundary of a 
polyhedron in $\mathbb{R}^3$.
This application, which motivated this study, is described in Section 2 of \cite{guiguesarxivcg2015},
following the lines of the seminal papers \cite{cornell}, \cite{mcguire}, which paved the way for PSHA.

In this paper, we describe  in Section \ref{algotriang1}
 an algorithm to compute
the CDF of $D$ when $S$ is a polygone using a triangulation
of the polygone.
This amounts to computing the area of the intersection of a disk 
and a triangle. To solve this problem, we enumerate all possible
shapes for this intersection, identify in which of these cases
we are (using tests depending on the disk and triangle considered),
and compute the area of this shape. This shape can be decomposed
as a union of triangles and lenses and therefore its area
can be easily computed analytically.
Finally, in Section \ref{lib}, we describe the main functions
of an open source library, available at \url{https://github.com/vguigues/Areas_Library},
implementing the algorithms from \cite{guiguesarxivcg2015} and 
Section \ref{algotriang1}. 

Throughout the paper, we use the following notation. 
For a point $A$ in $\mathbb{R}^{2}$,
we denote its coordinates with respect to a given Cartesian coordinate system by $x_A$ and  $y_A$.
For two points $A, B \in \mathbb{R}^{2}$,
$\overline{AB}$ is the line segment 
joining points $A$ and $B$, i.e., 
$\overline{AB}=\{tA + (1-t)B \;:\;t \in [0,1]\}$,
$(AB)=\{tA + (1-t)B \;:\;t \in \mathbb{R}\}$ is the line passing through
$A$ and $B$, and  
$\overrightarrow{AB}$ is the vector whose
coordinates are $(x_B-x_A, y_B-y_A)$. Given two vectors $x, y \in \mathbb{R}^2$, we denote
the usual scalar product of $x$ and $y$ in $\mathbb{R}^2$ by $\langle x, y \rangle = x^\transp y$.
For $P \in \mathbb{R}^2$, we denote the circle and the disk of center $P$ and radius $R$
by respectively $\mathcal{C}(P,R)$ and $\mathcal{D}(P,R)$.
 
\section{An algorithm based on a triangulation of the polygone}\label{algotriang1}

Let $S$ be a simple polygone contained in a plane
given by its extremal points $\{S_1, S_2, \ldots, S_n\}$
where the boundary of $S$ is $\bigcup_{i=1}^{n} \overline{S_i S_{i+1}}$
where $S_i \neq S_j$ for $i \neq j$
and $1 \leq i,j \leq n$.

The computation, at a given value $R$, of the CDF of the distance between a given point
$P$ and $S$ requires computing the area of the intersection
of $S$ and 
of the ball of center
$P$ and radius $R$.
Without loss generality, we will assume that $P$ is in the plane
containing $S$ (if this is not the case, we can project
$P$ onto this plane and modify the value of the radius, see 
\cite{guiguesarxivcg2015}} for details).
The area of the intersection can be obtained 
computing a
 triangulation of the polyhedron
(with complexity $n \log(n)$), then computing the area
of intersection of disk $\mathcal{D}(P,R)$ with the $n-2$ triangles of the triangulation,
and summing these $n-2$ areas.
Therefore, we need to devise an algorithm to compute the area of the intersection of 
a disk and a triangle and we proceed as follows.

Consider a triangle with vertices $A, B, C$ and a disk
of center $P$ and radius $R$. We want to compute the area
of the intesection of this triangle and this disk.

\begin{figure}[H]
\centering
\includegraphics[scale=0.4]{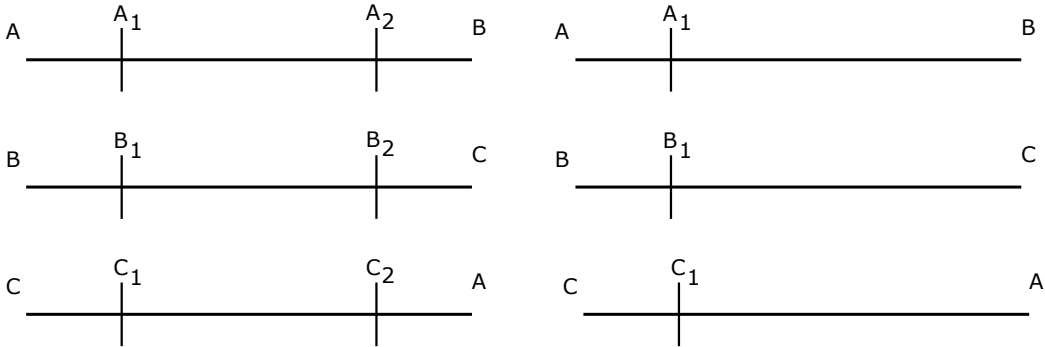}
\caption{Intersections between the sides
of the triangle and the cercle $\mathcal{C}(P,R)$.} 
\label{figinter}
\end{figure}

We first compute:
\begin{itemize}
\item the $n_1$ intersections of 
$\overline{AB}$
and $\mathcal{C}(P,R)$ denoted by $A_1$ when
$n_1=1$ and $A_1, A_2$ when $n_1=2$ where 
$\|\overrightarrow{A A_1}\|_2 < \|\overrightarrow{A A_2}\|_2$
(see Figure \ref{figinter});
\item the $n_2$ intersections of 
$\overline{BC}$
and $\mathcal{C}(P,R)$ denoted by $B_1$ when
$n_2=1$ and $B_1, B_2$ when $n_2=2$ where 
$\|\overrightarrow{B B_1}\|_2 < \|\overrightarrow{B B_2}\|_2$
(see Figure \ref{figinter});
\item the $n_3$ intersections of 
$\overline{AC}$
and $\mathcal{C}(P,R)$ denoted by $C_1$ when
$n_3=1$ and $C_1, C_2$ when $n_3=2$ where 
$\|\overrightarrow{C C_1}\|_2 < \|\overrightarrow{C C_2}\|_2$
(see Figure \ref{figinter}).
\end{itemize}
Each $n_i, i=1,2,3$, can take 3 values ($0, 1$, or $2$)
and therefore the triple $(n_1, n_2, n_3)$ can take
27 possible different values.
We associate to base 3 number $n_3 n_2 n_1$ its decimal
equivalent, denoted by {\tt{Code}} in what follows,
and given by ${\tt{Code}}=9n_3 + 3 n_2 +n_1$.
To identify the shape of the intersection between the triangle
and the disk, we branch according to the value of
variable {\tt{Code}} and for a given value of {\tt{Code}},
we consider all possible shapes for the intersection of the triangle
and the disk. Obviously, for
all values of {\tt{Code}} corresponding
to different permutations of the same triple $(n_1,n_2,n_3)$
the different possible shapes for the intersection are the same.
We now discuss for every possible value of variable 
{\tt{Code}} how to compute the area of the intersection.\\

\par $\bullet \; \; {\textbf{Code=0.}}$\\
\par
\begin{figure}[H]
\centering
\begin{tabular}{ccc}
\includegraphics[scale=0.3]{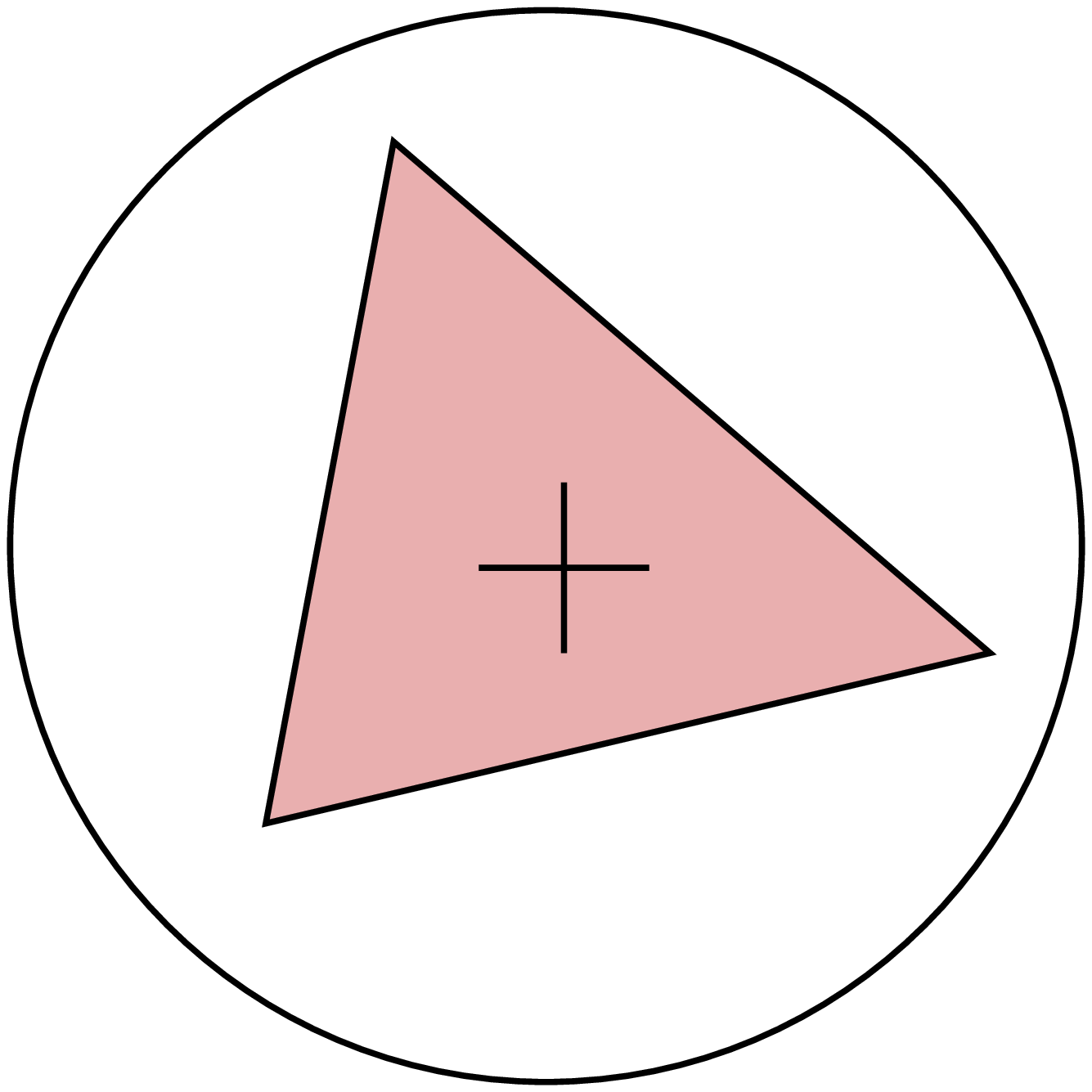}
&
\includegraphics[scale=0.3]{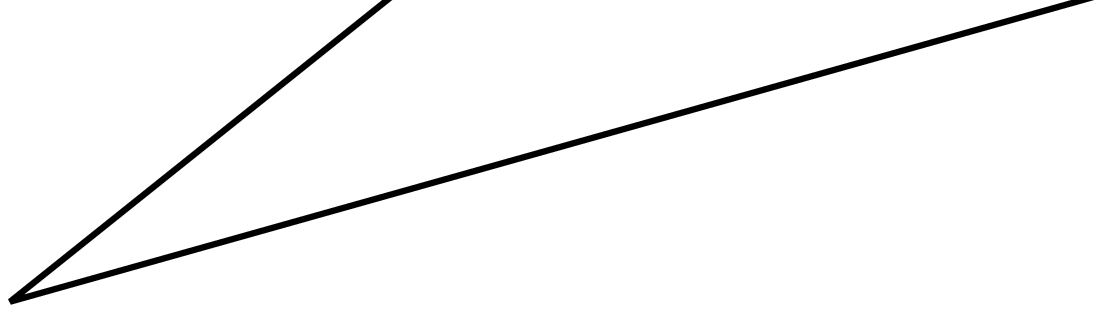}
&
\includegraphics[scale=0.3]{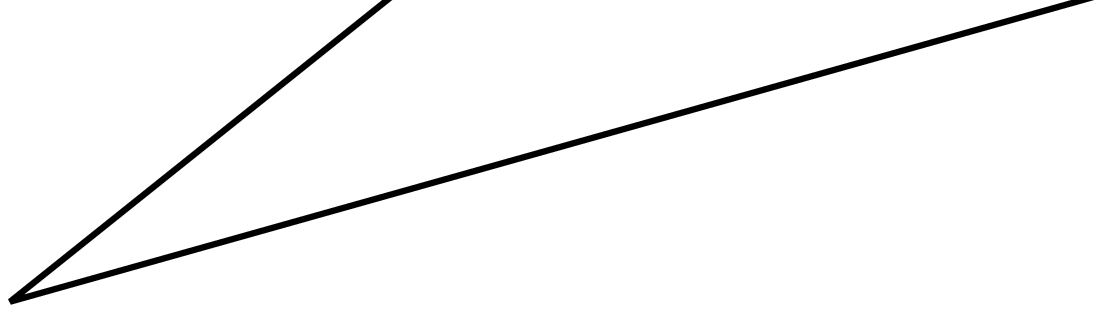}\\
(a)  & (b) & (c)\\
\end{tabular}
\caption{Possible shapes for the intersection when ${\tt{Code}}=0$.} 
\label{fig000}
\end{figure}

 In this case, there is no intersection
between the circle $\mathcal{C}(P,d)$ and the triangle.
There are three possible shapes for the intersection represented
in Figure \ref{fig000}: (a) the triangle is contained in the disk,
(b) the disk is outside the triangle, and (c) the disk is inside
the triangle. 
To know which of subcases (a), (b), or (c)
we are in, we compute  
\begin{equation}\label{defdadbdc}
d_A = \|\overrightarrow{PA}\|_2,
 d_B = \|\overrightarrow{PB}\|_2, d_C = \|\overrightarrow{PC}\|_2.
\end{equation}
In what follows, we denote by 
\begin{equation}\label{areatriangle}
T(A,B,C)= 
\frac{1}{2}\Big|x_A(y_B-y_C)+x_B (y_C -y_A) +x_C( y_A-y_B)\Big|
\end{equation}
the area of triangle $ABC$.
If $d_A<R, d_B<R$, and $d_C<R$ then 
the area is $T(A,B,C)$. Otherwise 
either $P$ is inside the triangle and the area
is $\pi R^2$ or it is outside and the area is null.
To know if $P$ is inside or outside the triangle,
we compute the crossing number for $P$ and the triangle
(see \cite{rourkebook} for the definition of the crossing number
and for instance \cite{guiguesarxivcg2015, rourkebook} for an algorithm to compute it) and the minimal distance $d_{\min}$ from $P$ to the border
of the triangle. 
Knowing that the crossing number is odd if and only if
$P$ belongs to the relative interior of the triangle,
$P$ is inside the triangle
if and only if $d_{\min}=0$ or the crossing
number is odd.\\
\par $\bullet \; {\textbf{Code}}=1, 3, 9$, corresponding
to $(n_1,n_2,n_3) \in \{(1,0,0),(0,1,0),(0,0,1)\}$, i.e., one side
of the triangle has a single intersection with the circle
and the remaining two have no intersection.\\

\par
\begin{figure}[H]
\centering
\begin{tabular}{cc}
\includegraphics[scale=0.32]{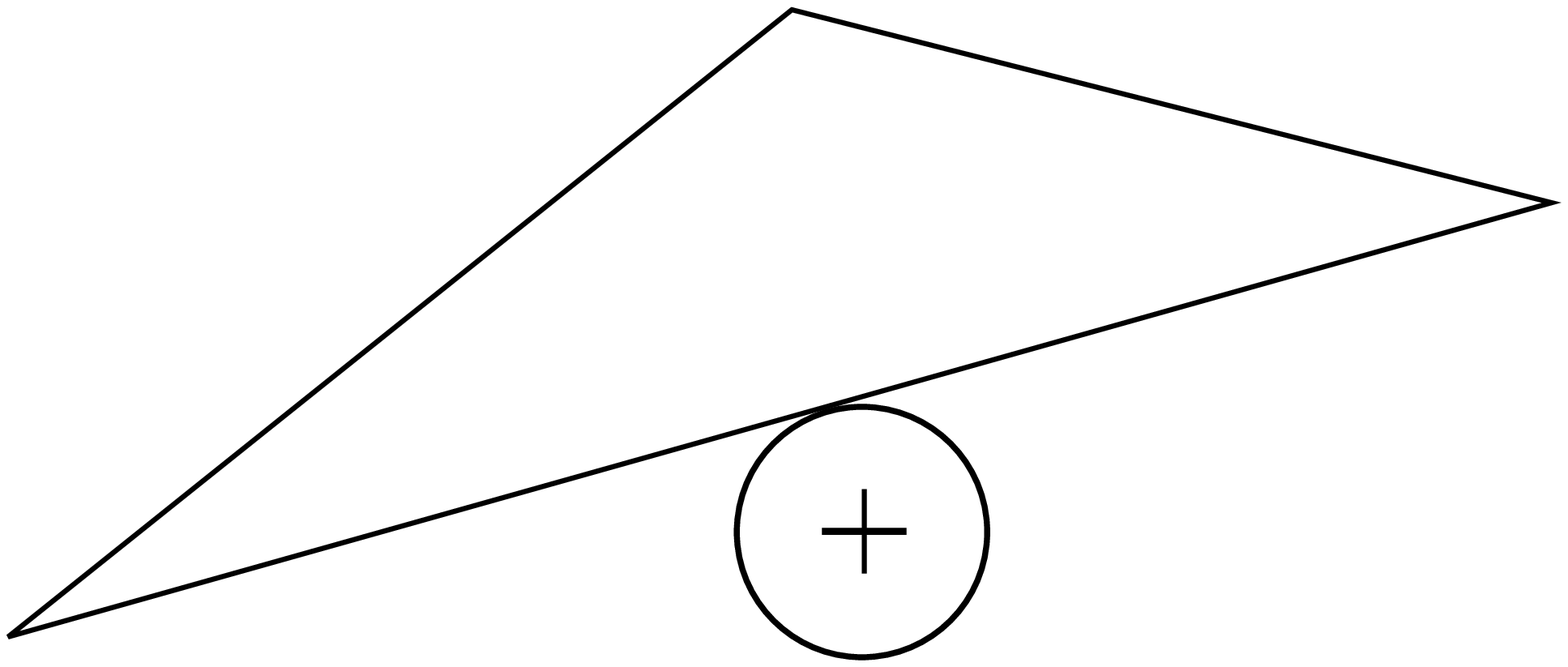}
&
\includegraphics[scale=0.32]{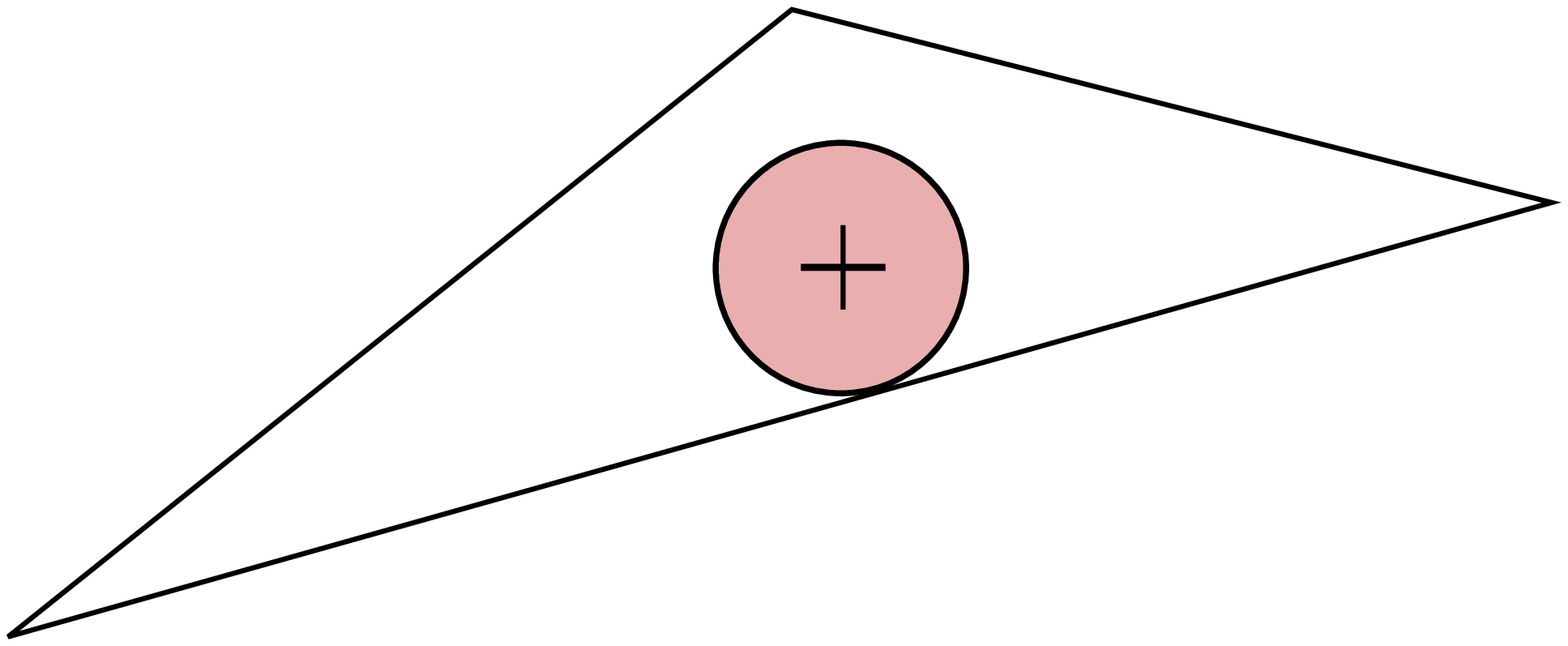}\\
(a)  & (b)\\
\end{tabular}
\caption{Possible shapes for the intersection when {\tt{Code=$1,3,9$}}.} 
\label{fig001}
\end{figure}

 There are two possible shapes for the intersection represented 
in Figure \ref{fig001}-(a),(b).
In case (a), $P$ is outside the triangle and the area is null while in
case (b), $P$ is inside the triangle and the area is $\pi R^2$. We have already seen how 
to differentiate these two cases on the basis of $d_{\min}$ and of the
crossing number.\\ 
\par $\bullet \; {\textbf{Code=2, 6, 18}},$ obtained when $(n_1,n_2,n_3)$ is
$(2,0,0), (0,2,0), (0,0,2)$, respectively.\\
\par
\begin{figure}[H]
\centering
\begin{tabular}{cc}
\includegraphics[scale=0.32]{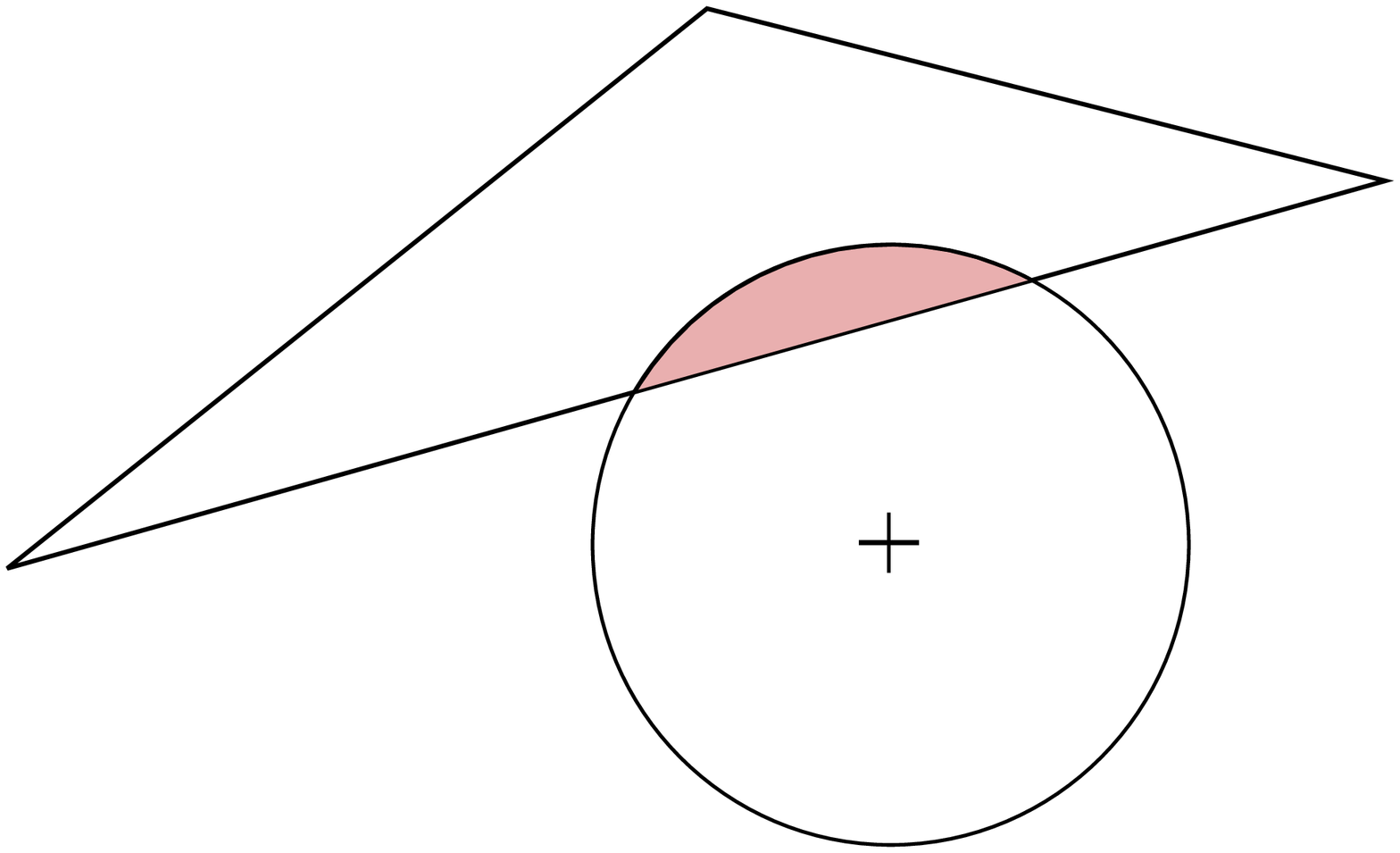}
&
\includegraphics[scale=0.32]{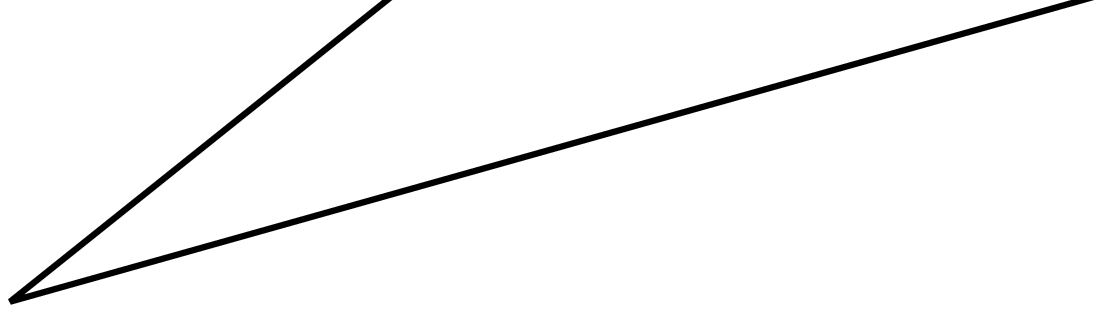}\\
(a)  & (b) \\
\end{tabular}
\caption{{Possible shapes for the intersection when \tt{Code=$2,6, 18$.}}}
\label{fig002}
\end{figure}

 There are two possible shapes for the intersection represented
in Figure \ref{fig002}-(a), (b).
In each case, the intersection is a lens, of area 
$\leq 0.5 \pi R^2$ in case (a) and $\geq 0.5 \pi R^2$
in case (b). Let us recall how to compute analytically
these areas.

\begin{figure}[H]
\centering
\begin{tabular}{ccc}
\includegraphics[scale=0.25]{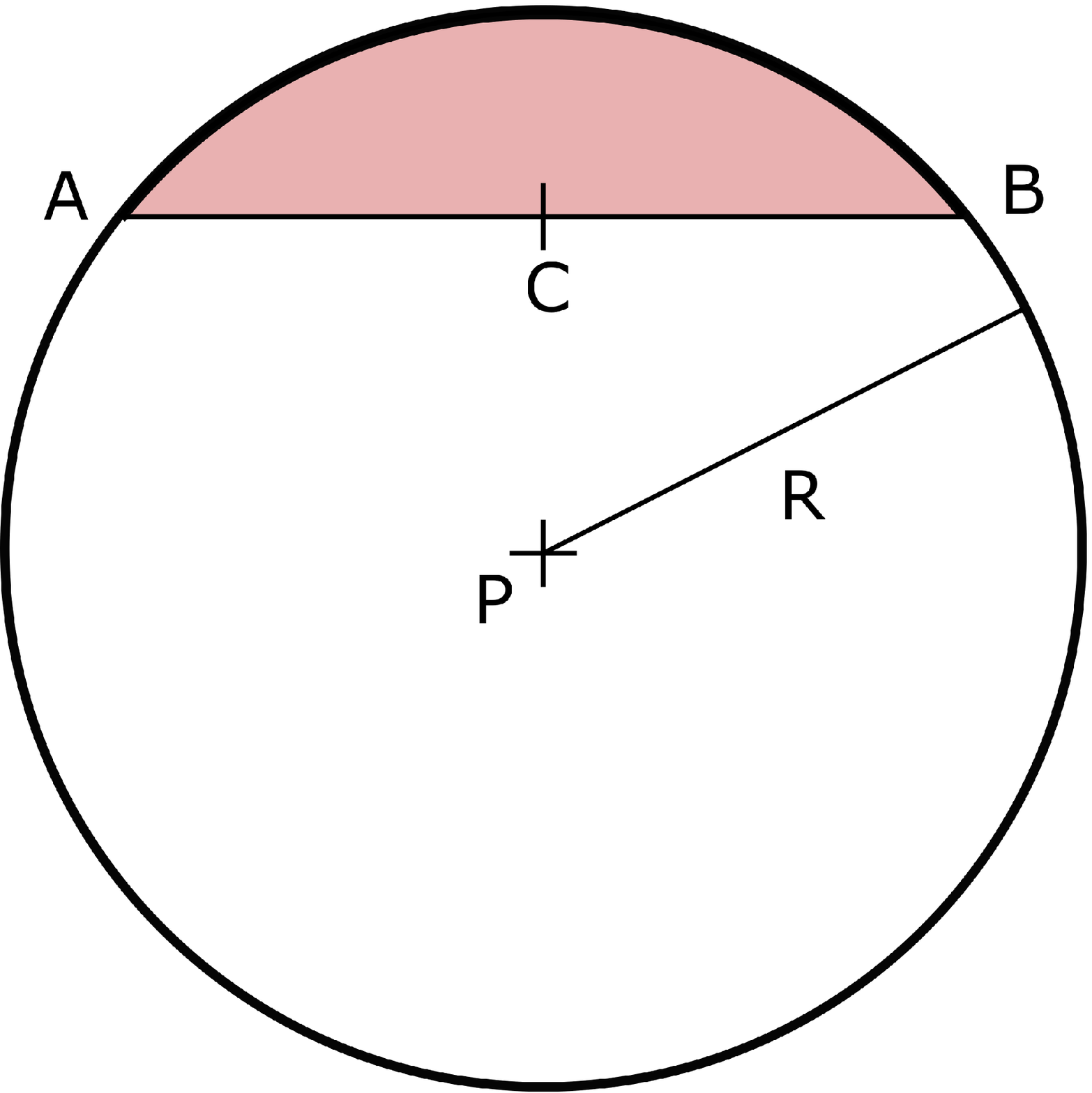}
&\hspace*{1cm}&
\includegraphics[scale=0.25]{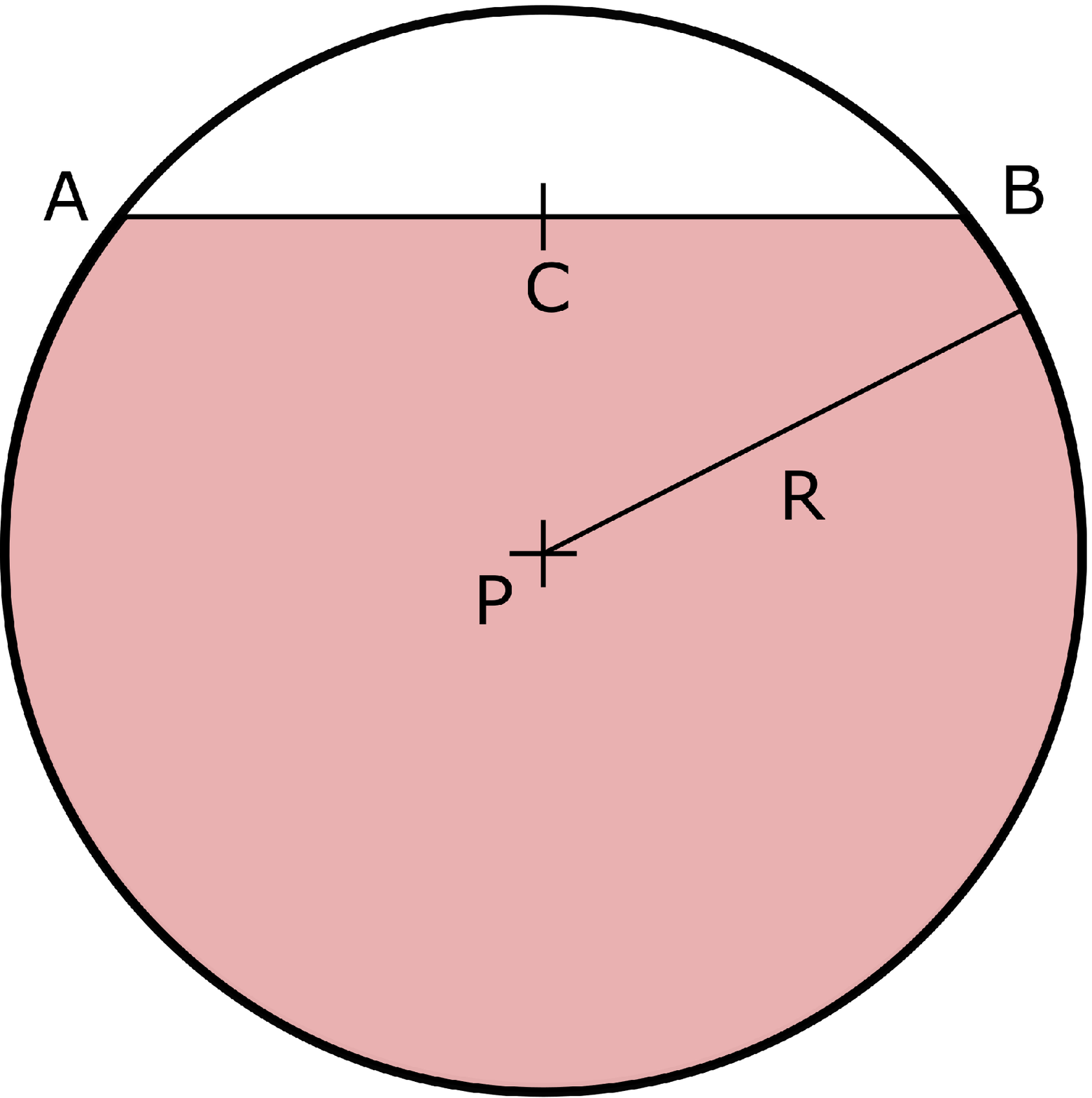}
\\
(a)&  & (b)\\
\end{tabular}
\caption{Two lenses given by chord $\overline{AB}$ in a disk of center $P$ and radius $R$.} 
\label{figlens}
\end{figure}

Consider a chord $\overline{AB}$ of circle $\mathcal{C}(P,R)$.
It defines two lenses represented in Figure \ref{figlens}-(a), (b).
Let us recall the formula for the area 
$L(P,R,A,B)$ of the lens in case (a) (in case (b), it is given by 
$\pi R^2 - L(P,R,A,B)$).
In case (a), let $C=\frac{A+B}{2}$ and
let $c\theta$ be the cosine of acute angle
$\angle CPB$ given by
$$
c\theta = \frac{
\langle \overrightarrow{PC} , \overrightarrow{PB} \rangle}{R \| \overrightarrow{PC}  \|_2}.
$$
Then 
$$
L(P,R,A,B)=R^2 \arccos( c \theta ) -R^2 c\theta \sqrt{1-c\theta^2}.
$$
With this notation, when ${\tt{Code}}=18$, i.e., when 
$(n_3,n_2,n_1)=(2,0,0)$, in case (a), the area of the intersection is
given by $L(P,R,C_1,C_2)$.
The cases where ${\tt{Code}}=2, 6$ are dealt with by appropriate permutation
of the intersection points.\\
\par $\bullet \; {\textbf{Code=4, 10, 12,}}$ obtained when 
$(n_3,n_2,n_1)$ is $(0,1,1)$, $(1,0,1)$, $(1,1,0)$.\\
\par
\begin{figure}
\centering
\begin{tabular}{cc}
\includegraphics[scale=0.35]{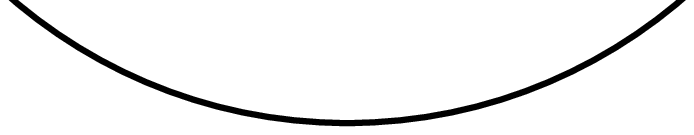}
&
\includegraphics[scale=0.35]{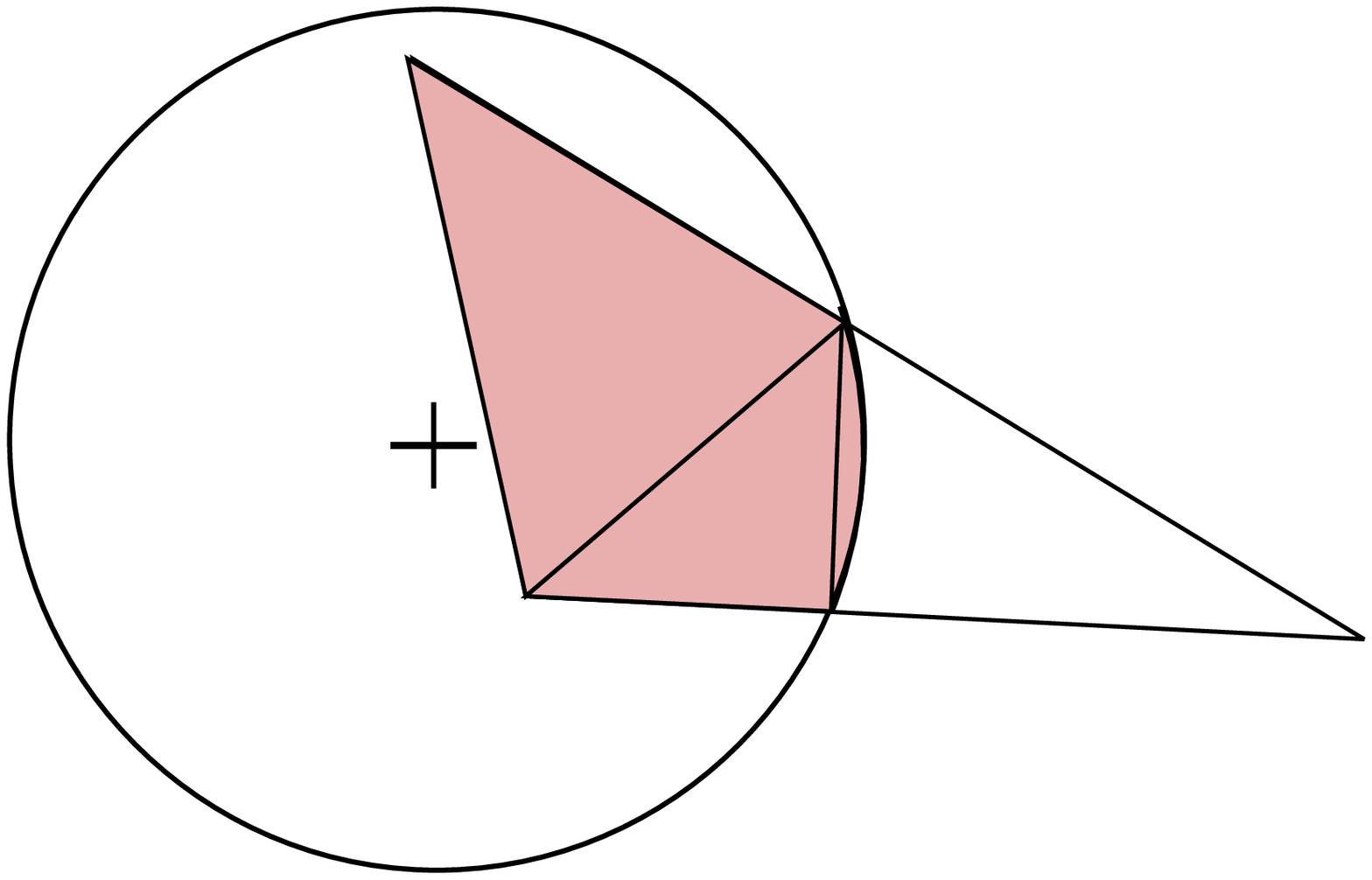}\\
(a)  & (b) \\
\includegraphics[scale=0.35]{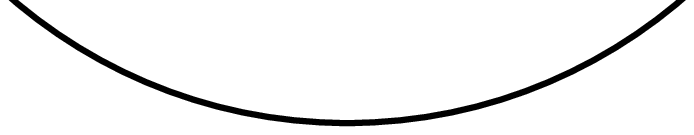}
&
\includegraphics[scale=0.35]{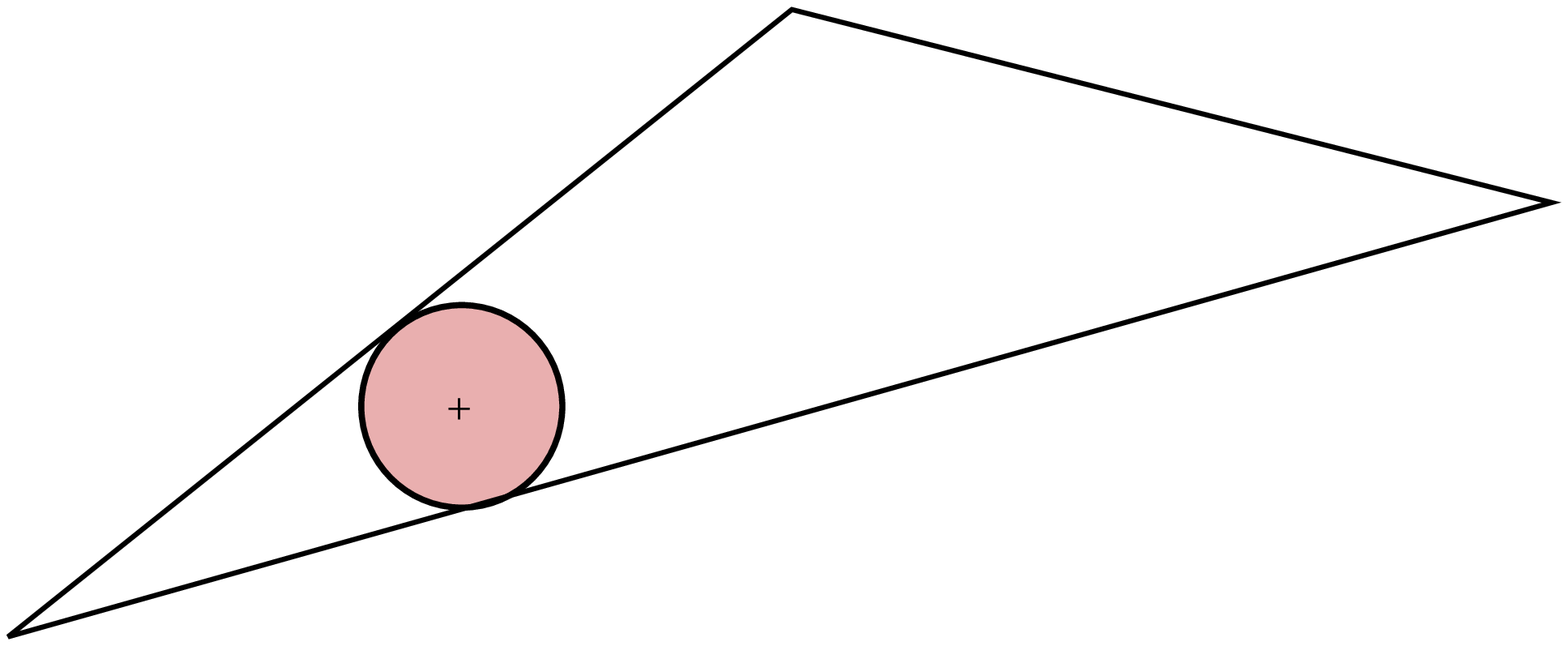}\\
(c)  & (d) \\
\includegraphics[scale=0.35]{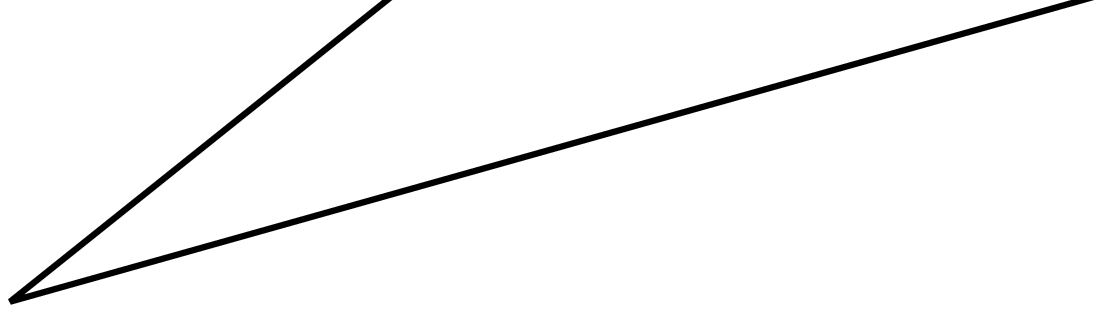}
&
\includegraphics[scale=0.35]{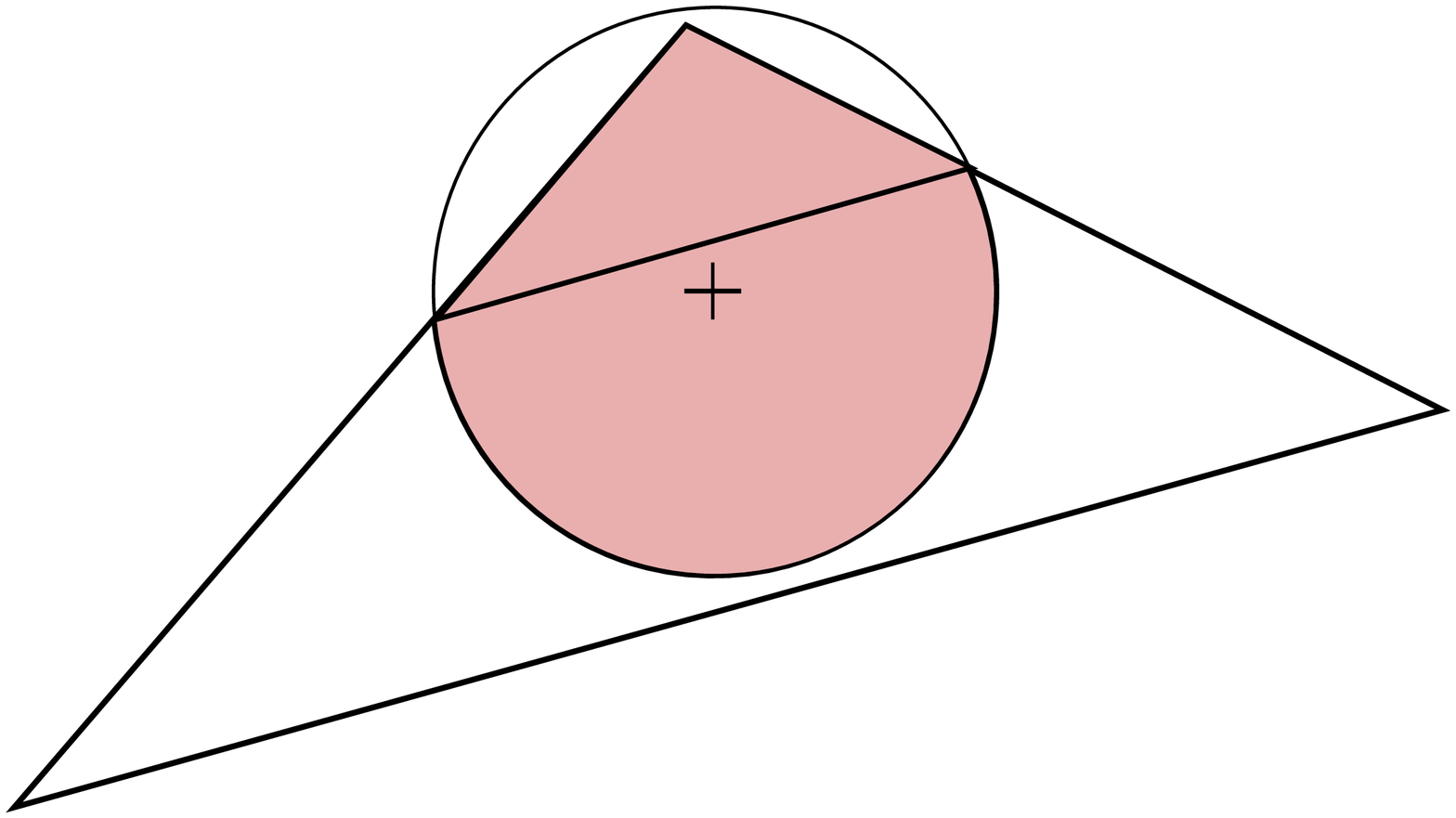}\\
(e)  & (f) \\
\end{tabular}
\caption{Possible shapes for the intersection when {\tt{Code=$4,10$, or $12$}}.} 
\label{fig011}
\end{figure}

 The possible shapes for the intersection between the triangle and 
disk are given in Figure \ref{fig011}.
We discuss how to compute the intersection area when 
$(n_3, n_2, n_1)=(1,1,0)$, i.e., ${\tt{Code=12}}$. The cases 
{\tt{Code=4,10}},
are similar, obtained by appropriate permutation of $A,B,C$
and the intersection points. 

Let us first discuss on how to distinguish between cases (e) and (f).
In theses cases, the intersection is the union of a triangle and a lens.
In case (e) the lens has area lower than or equal to $0.5 \pi R^2$
while in case (f) the lens has area greater than $0.5 \pi R^2$.
In case (f), $P$ and $C$ are in two different half-spaces separated
by line $(C_1 B_1)$. More precisely, let us introduce the function
\begin{center}
{\tt{[out]}}={\tt{Are\_In}}\_{\tt{Same}}\_{\tt{Half}}\_{\tt{Space}}(C,P,A,B)
\end{center}
with inputs four points $C,P,A,B$ in $\mathbb{R}^2$ which returns 
0 if $C$ and $P$ are in two different half-spaces separated by line 
$(AB)$
and 1 otherwise. Clearly, {\tt{out}} is 1
if and only if ($x_A= x_B$ and $(x_A-x_P)(x_A-x_C) \geq 0$)
or ($x_A \neq x_B$ and
$(y_P - D_{A,B}(x_P))(y_C - D_{A,B}(x_C)) \geq 0$)
where
$$
D_{A,B}(x)=y_A + \frac{y_B-y_A}{x_B-x_A}(x-x_A).
$$
With this notation, the intersection area in cases (a)-(f) is
computed with the following pseudo-code where {\tt{area}}
will store the intersection area:\\

\par {\tt {area=0.}}
\par {\textbf{if }} $d_A<R$ and $d_B<R$ and $d_C=R$ then
$
{\tt{area}}=T(A,B,C),
$
\par {\textbf{else if }}$d_A<R$ and $d_B<R$,  and $d_C>R$ then
$
{\tt{area}}=L(P,R,C_1,B_1) + T(A,B,B_1) +  T(A,B_1,C_1),
$
\par {\textbf{else if }}$d_A>R$ and $d_B>R$ and $d_C>R$ then
$
{\tt{area}}=\pi R^2,
$
\par {\textbf{else if }} $d_A>R$ and $d_B>R$ and $d_C<R$ then
\par \hspace*{1cm}{\tt{[out]}}={\tt{Are\_In}}\_{\tt{Same}}\_{\tt{Half}}\_{\tt{Space}}$(C,P,B_1,C_1)$,
\par \hspace*{1cm}{\textbf{if }}out=1 then {\tt{area}}$=T(C,B_1,C_1)+L(P,R,B_1,C_1)$,
\par \hspace*{1cm}{\textbf{else }} {\tt{area}}$=T(C,B_1,C_1)+\pi R^2-L(P,R,B_1,C_1)$,
\par \hspace*{1cm}{\textbf{end if}}
\par {\textbf{end if}}\\

\par $\bullet \; {\textbf{Code=5,7,11,15,19,21,}}$
corresponding to $(n_3,n_2,n_1) \in \{(0,1,2)$, 
$(0,2,1)$,$(1,0,2)$, $(1,2,0)$, $(2,0,1)$,
$(2,1,0)\}$.\\
\par
\begin{figure}
\centering
\begin{tabular}{cc}
\includegraphics[scale=0.3]{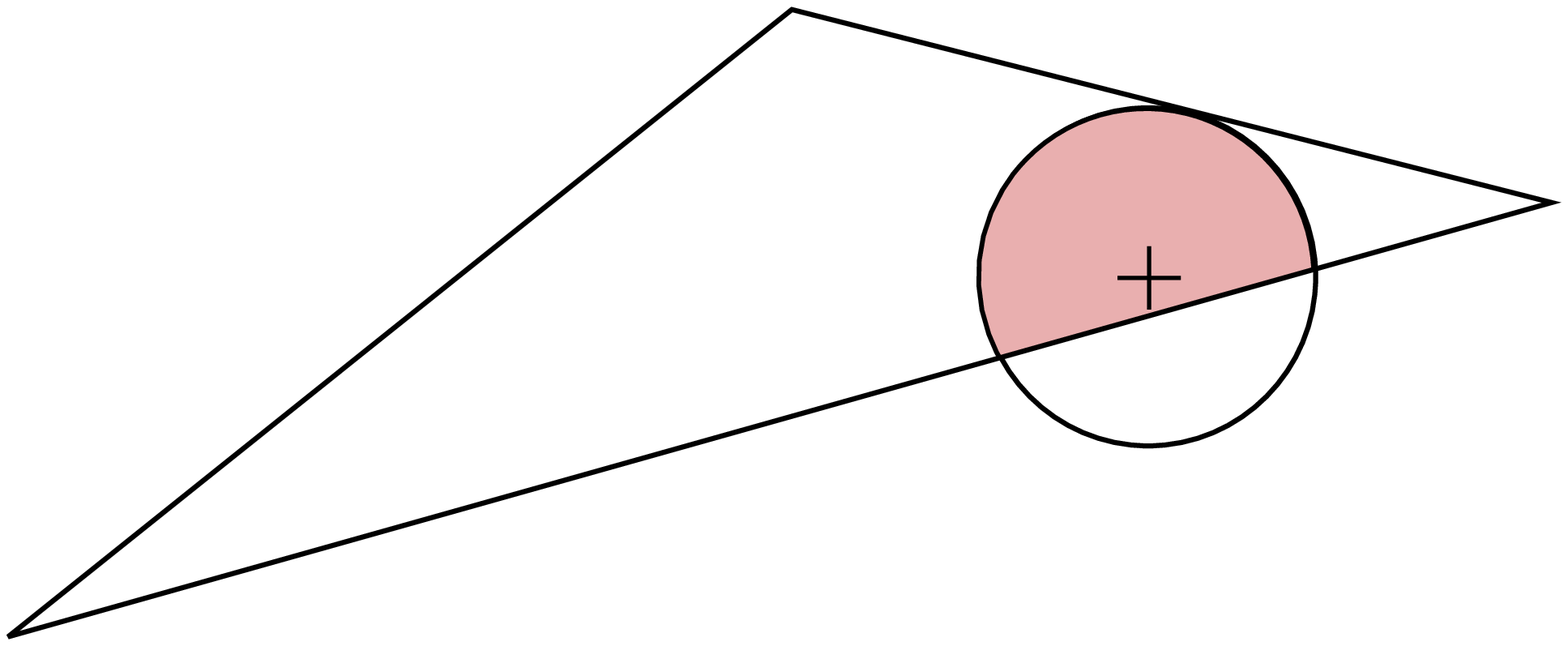}
&
\includegraphics[scale=0.3]{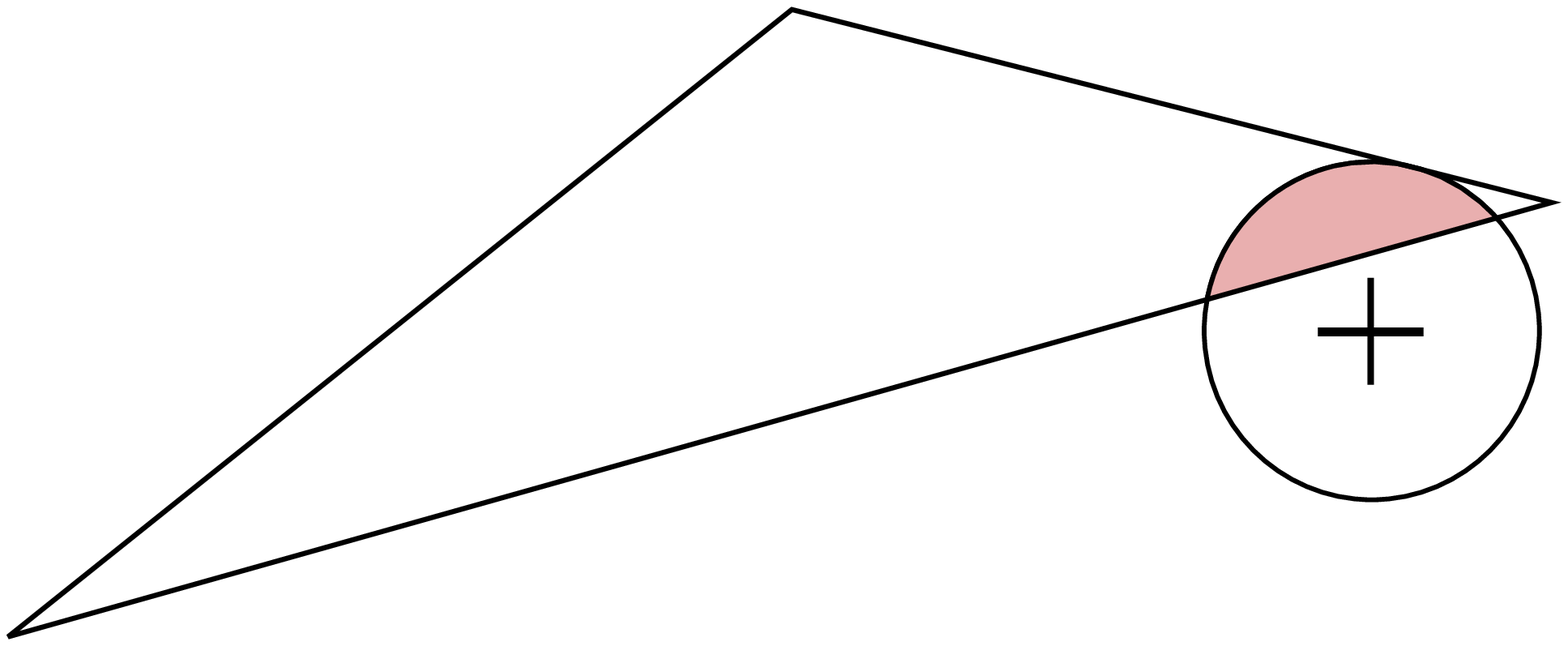}\\
(a)  & (b) \\
\includegraphics[scale=0.3]{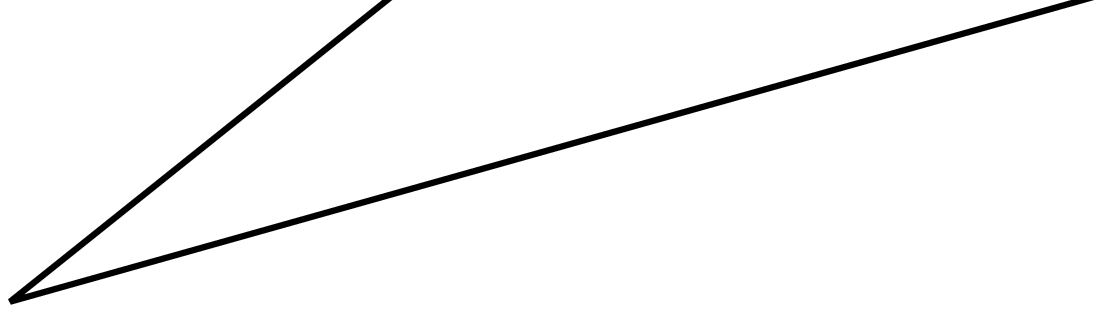}
&
\includegraphics[scale=0.3]{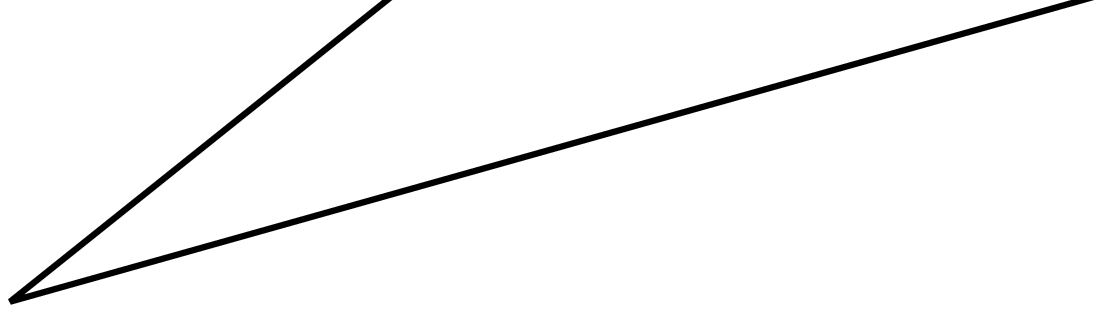}\\
(c)  & (d) \\
\end{tabular}
\caption{Possible shapes for the intersection when {\tt{Code=$5,7, 11, 15, 19$, or $21$}}.} 
\label{fig201}
\end{figure}

 The possible shapes for the intersection are represented
in Figure \ref{fig201}.
It follows that 
when {\tt{Code}=5} or 11, i.e., when
$(n_3,n_2,n_1)$ is
$(0,1,2)$ or $(1,0,2)$, the area of the intersection
is computed with the following pseudo-code (stored
in variable {\tt{area}}):\\
\par {\textbf{if }} the crossing number for $P$ and the
triangle is odd or $d_{\min}=0$ then 
{\tt{area}}$=\pi R^2 - L(P,R,A_1,A_2)$
\par {\textbf{else }}{\tt{area}}=$L(P,R,A_1,A_2)$.
\par {\textbf{end if}}\\
\par The pseudo-codes when {\tt{Code}=7,15,19,21},
are obtained by appropriate permutations of
the intersection points.\\
\par $\bullet \; {\textbf{Code=8, 20, 24}}$
obtained when $(n_3,n_2,n_1)=(0,2,2)$, $(2,0,2)$, $(2,2,0)$.\\
\par
\begin{figure}
\centering
\begin{tabular}{ccc}
\includegraphics[scale=0.35]{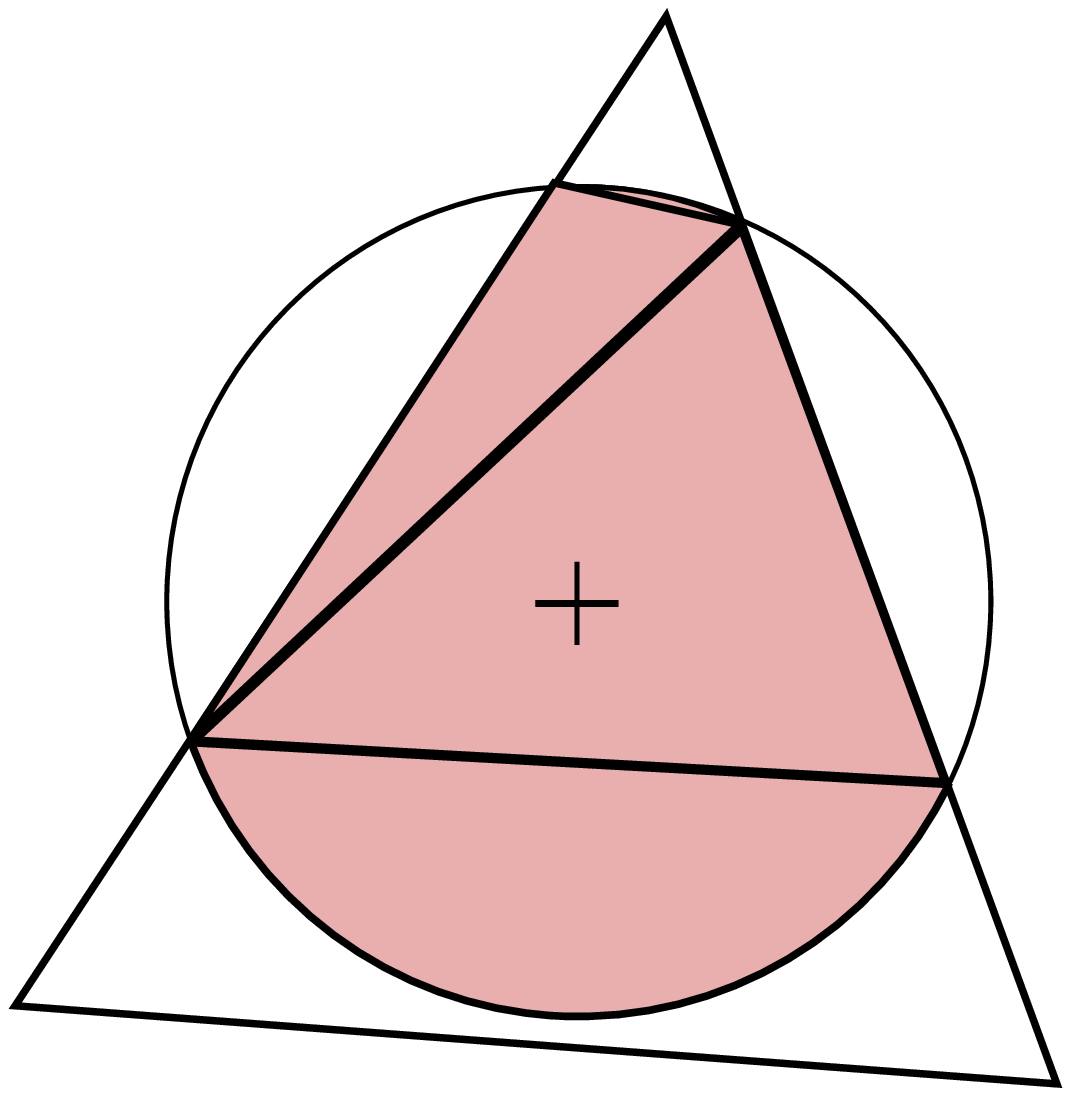}
&&
\includegraphics[scale=0.35]{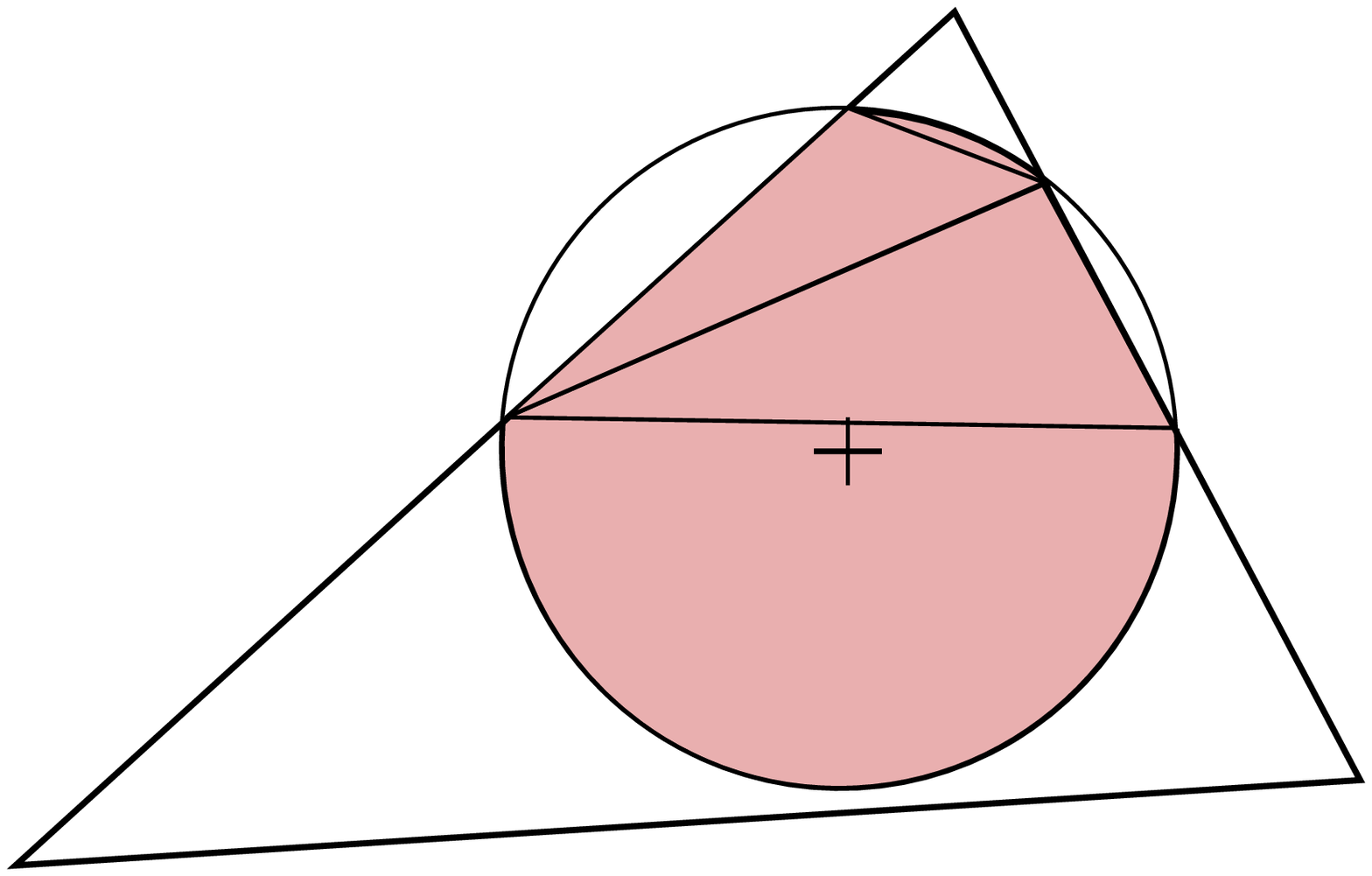}\\
(a)  & &(b)\\
\includegraphics[scale=0.35]{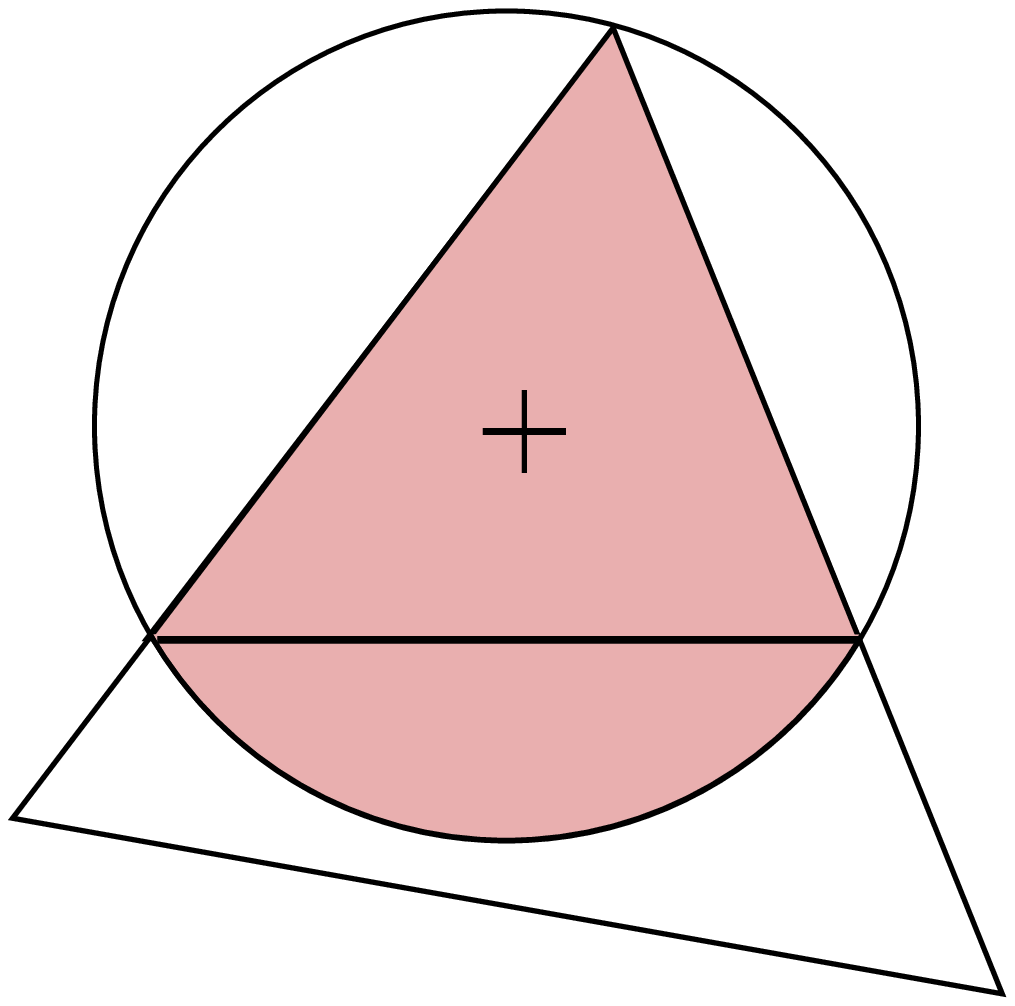}
&&
\includegraphics[scale=0.35]{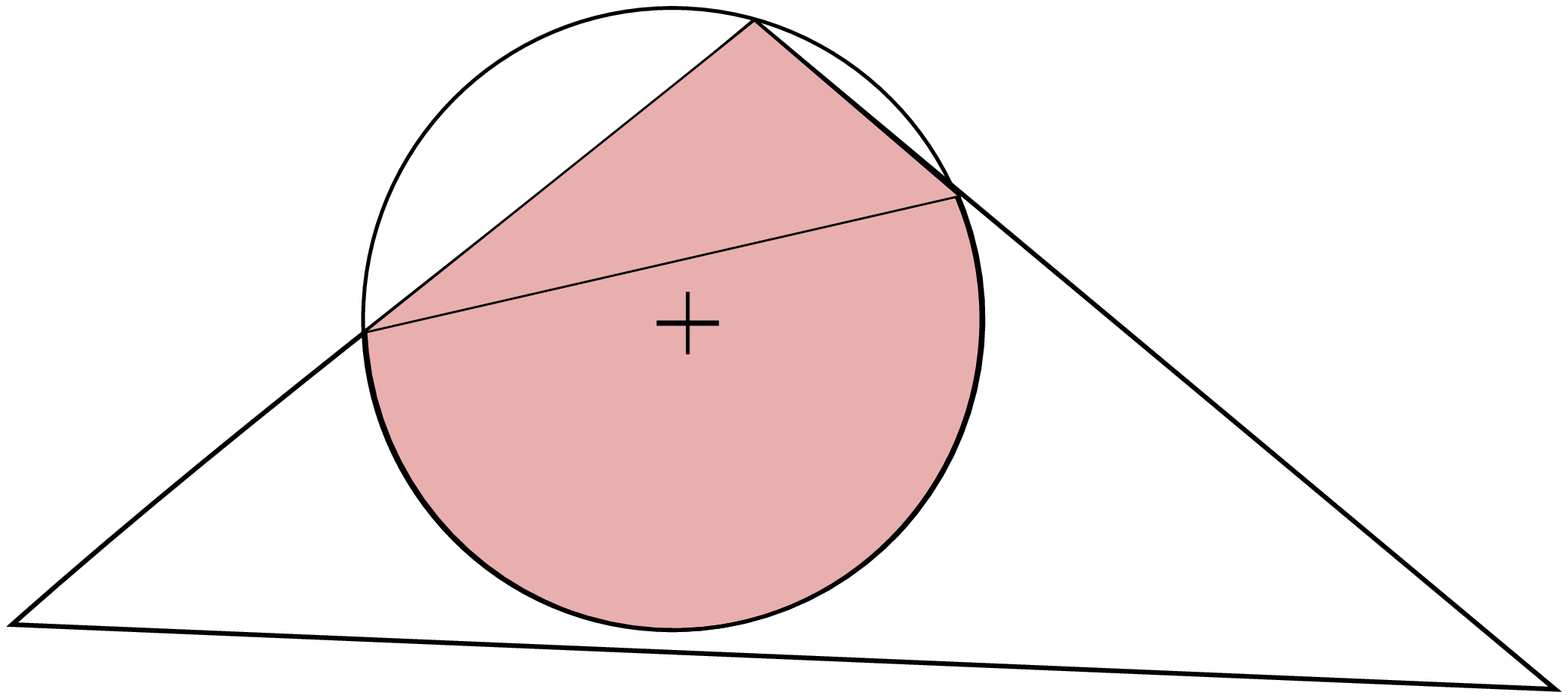}\\
(c)&  & (d) \\
\end{tabular}
\caption{Possible shapes for the intersection when {\tt{Code=$8, 20$, or $24$}}} 
\label{fig220}
\end{figure}

 The possible shapes for the intersection are represented
in Figure \ref{fig220}.
It follows that when {\tt{Code=8}}, i.e., when
$(n_3,n_2,n_1)=(0,2,2)$ the pseudo-code for computing the
area of the intersection (stored in variable {\tt{area}})
is the following:\\

\par {\tt{area=0}}.
\par {\tt{[out]}}={\tt{Are\_In}}\_{\tt{Same}}\_{\tt{Half}}\_{\tt{Space}}$(B,P,A_1,B_2)$
\par {\textbf{if}} $d_B>R$
\par \hspace*{0.5cm}{\textbf{if}} {\tt{out}}=1
\par \hspace*{0.8cm}{\tt{area}}=
$L(P,R,A_1,B_2)+T(B_1,A_1,B_2)+T(A_2,A_1,B_1)+L(P,R,A_2,B_1)$,
\par \hspace*{0.5cm}{\textbf{else}}
\par \hspace*{0.8cm}{\tt{area}}=
$\pi R^2 - L(P,R,A_1,B_2)+T(B_1,A_1,B_2)+T(A_2,A_1,B_1)+L(P,R,A_2,B_1)$,
\par \hspace*{0.5cm}{\textbf{end if}}
\par {\textbf{else}}
\par \hspace*{0.5cm}{\textbf{if}} {\tt{out}}=1
\par \hspace*{0.8cm}{\tt{area}}=
$L(P,R,A_1,B_2)+T(B,A_1,B_2)$,
\par \hspace*{0.5cm}{\textbf{else}}
\par \hspace*{0.8cm}{\tt{area}}=
$\pi R^2 - L(P,R,A_1,B_2)+T(B,A_1,B_2)$.
\par \hspace*{0.5cm}{\textbf{end if}}
\par {\textbf{end if}}\\
The pseudo-codes for
{\tt{Code}}$=20$ and $24$
are obtained by appropriate permutations of
$A,B,C$, and the intersection points.\\

\par $\bullet \; {\textbf{Code=13}}$ obtained when
$(n_3,n_2,n_1)=(1,1,1)$.\\

\par
\begin{figure}
\centering
\begin{tabular}{ccc}
\includegraphics[scale=0.35]{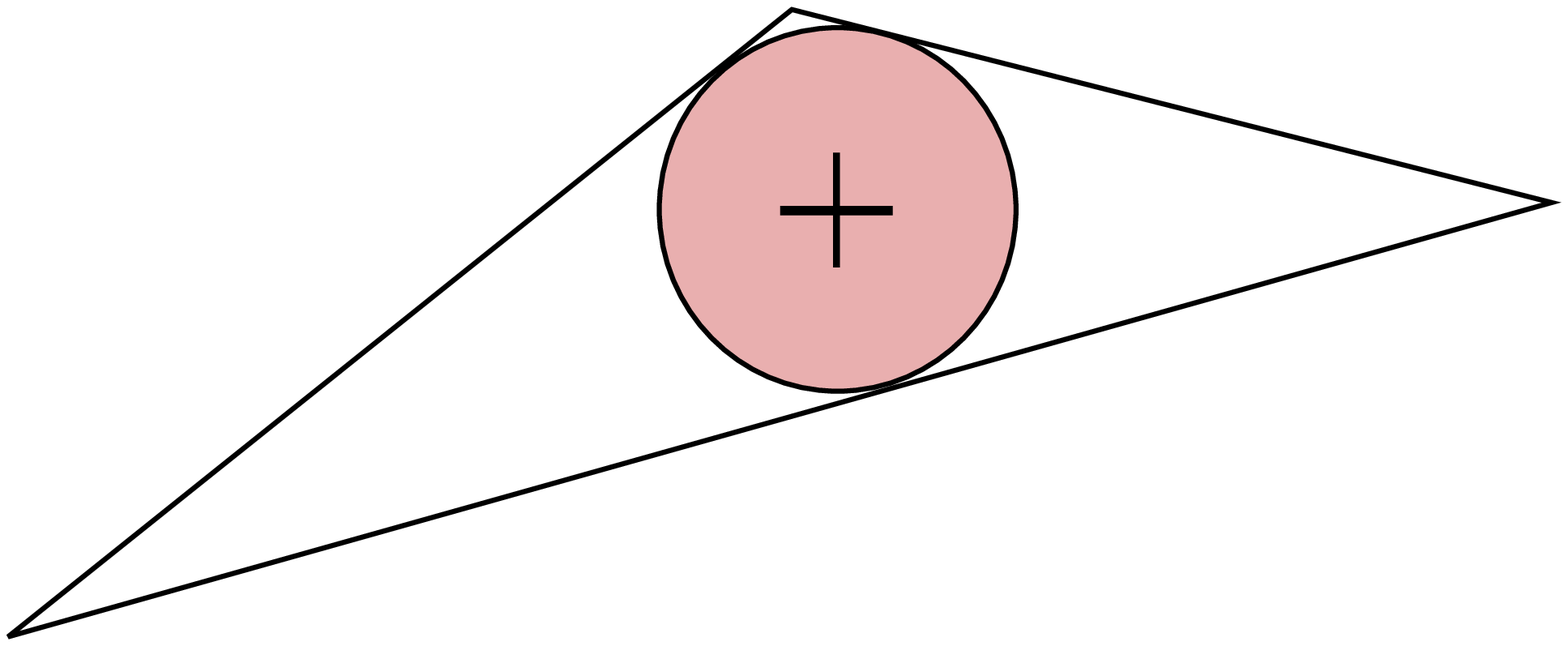}
&&
\includegraphics[scale=0.35]{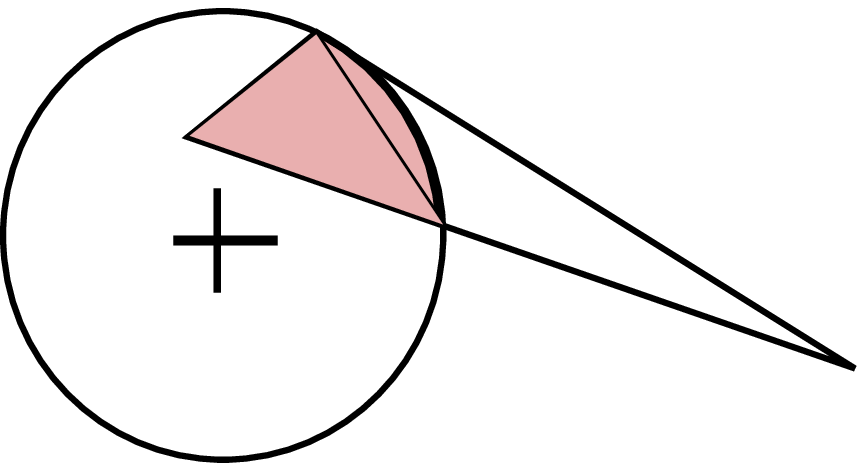}\\
(a)  & &(b)\\
\includegraphics[scale=0.35]{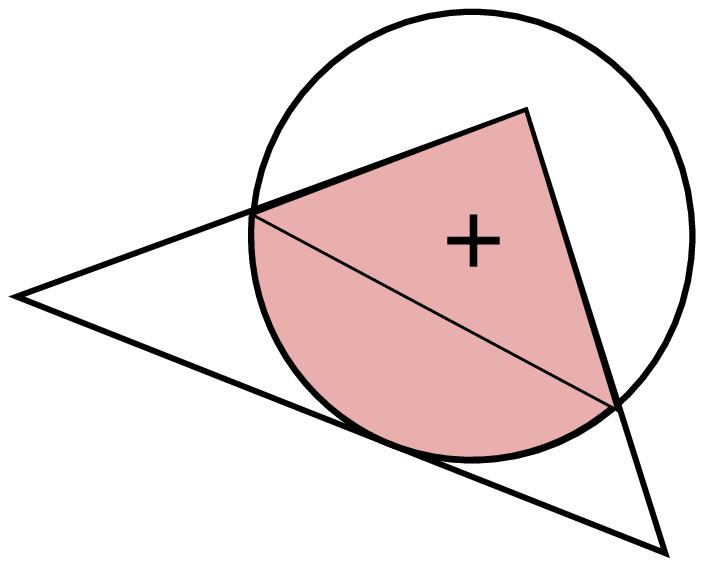}
&&
\includegraphics[scale=0.35]{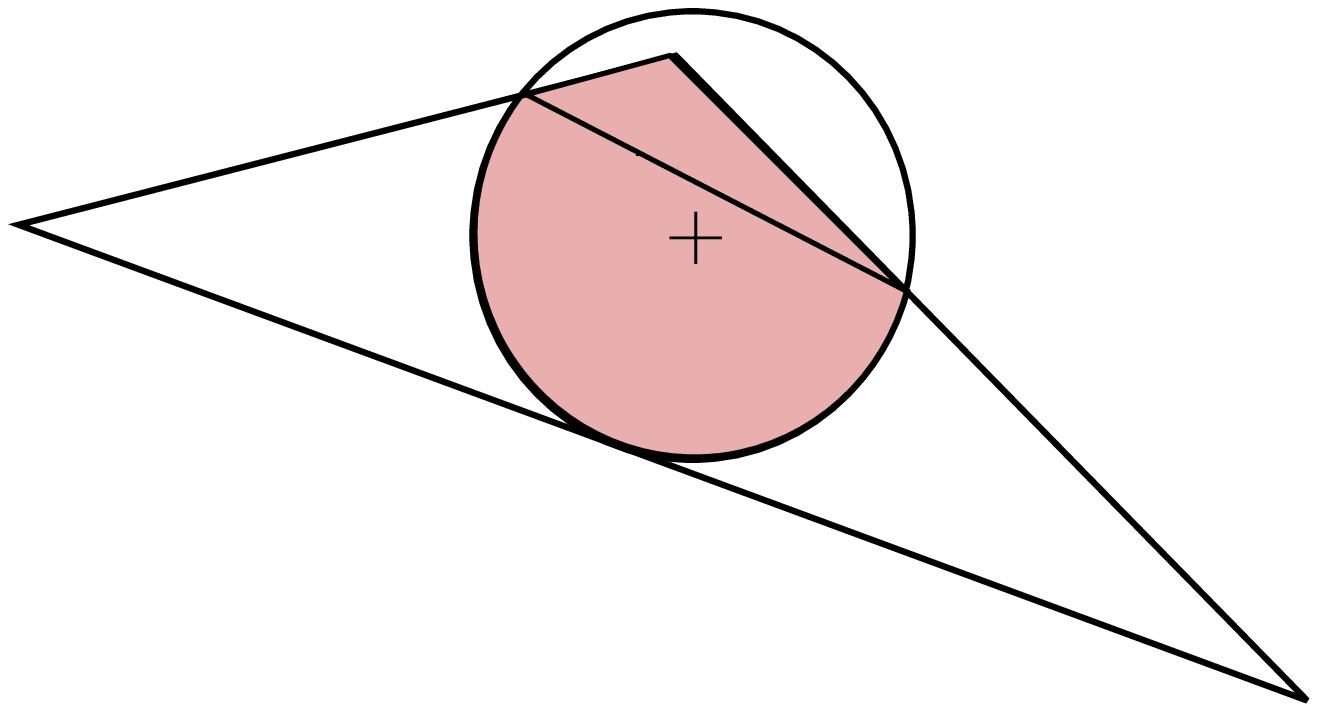}\\
(c) & & (d) \\
\end{tabular}
\caption{Possible shapes for the intersection when {\tt{Code=$13$}}.} 
\label{fig111}
\end{figure}

 The possible shapes for the intersection are
represented in Figure \ref{fig111}.
Therefore the pseudo-code for computing the intersection
area is the following (where 
the area of the intersection is stored  in variable {\tt{area}}):\\
\par {\textbf{if }}$d_A>R$ and $d_B>R$ and $d_C>R$ then
$
{\tt{area}}=\pi R^2,
$
\par {\textbf{else if }}$d_A=R$ and $d_B>R$ and $d_C<R$ then
$
{\tt{area}}=L(P,R,A,B_1)+T(A,C,B_1),
$
\par {\textbf{else if }}$d_A=R$ and $d_C>R$ and $d_B<R$ then
$
{\tt{area}}=L(P,R,A,B_1)+T(A,B,B_1),
$
\par {\textbf{else if }}$d_B=R$ and $d_A>R$ and $d_C<R$ then
$
{\tt{area}}=L(P,R,B,C_1)+T(C,B,C_1),
$
\par {\textbf{else if }}$d_B=R$ and $d_C>R$ and $d_A<R$ then
$
{\tt{area}}=L(P,R,B,C_1)+T(A,B,C_1),
$
\par {\textbf{else if }}$d_C=R$ and $d_B>R$ and $d_A<R$ then
$
{\tt{area}}=L(P,R,C,A_1)+T(C,A,A_1),
$
\par {\textbf{else if }}$d_C=R$ and $d_B<R$ and $d_A>R$ then
$
{\tt{area}}=L(P,R,C,A_1)+T(C,B,A_1),
$
\par {\textbf{else if }}$d_A<R$ and $d_B>R$ and $d_C>R$ then
\par \hspace*{0.8cm}{\tt{[out]}}={\tt{Are\_In}}\_{\tt{Same}}\_{\tt{Half}}\_{\tt{Space}}$(P,A,A_1,C_1)$,
\par \hspace*{0.8cm}{\textbf{if }}{\tt{out}=1} then {\tt{area}}$=L(P,R,A_1, C_1) + T(A, A_1, C_1)$,
\par \hspace*{0.8cm}{\textbf{else }}{\tt{area}}$=\pi R^2 - L(P,R,A_1, C_1) + T(A, A_1, C_1)$
\par \hspace*{0.8cm}{\textbf{end if}}

\par {\textbf{else if }}$d_B<R$ and $d_A>R$ and $d_C>R$ then
\par \hspace*{0.8cm}{\tt{[out]}}={\tt{Are\_In}}\_{\tt{Same}}\_{\tt{Half}}\_{\tt{Space}}$(P,B,A_1,B_1)$,
\par \hspace*{0.8cm}{\textbf{if }}{\tt{out}=1} then {\tt{area}}$=L(P,R,A_1,B_1) + T(B, A_1, B_1)$,
\par \hspace*{0.8cm}{\textbf{else }}{\tt{area}}$=\pi R^2 -L(P,R,A_1,B_1)+T(B, A_1, B_1)$
\par \hspace*{0.8cm}{\textbf{end if}}

\par {\textbf{else if }}$d_C<R$ and $d_A>R$ and $d_B>R$ then
\par \hspace*{0.8cm}{\tt{[out]}}={\tt{Are\_In}}\_{\tt{Same}}\_{\tt{Half}}\_{\tt{Space}}$(P,C,C_1,B_1)$,
\par \hspace*{0.8cm}{\textbf{if }}{\tt{out}=1} then {\tt{area}}$=L(P,R,C_1,B_1) + T(C,C_1,B_1)$,
\par \hspace*{0.8cm}{\textbf{else }}{\tt{area}}$=\pi R^2 -L(P,R,C_1,B_1) + T(C,C_1,B_1)$,
\par \hspace*{0.8cm}{\textbf{end if}}
\par {\textbf{end if}}\\

\par
\begin{figure}
\centering
\begin{tabular}{c}
\begin{tabular}{ccc}
\includegraphics[scale=0.5]{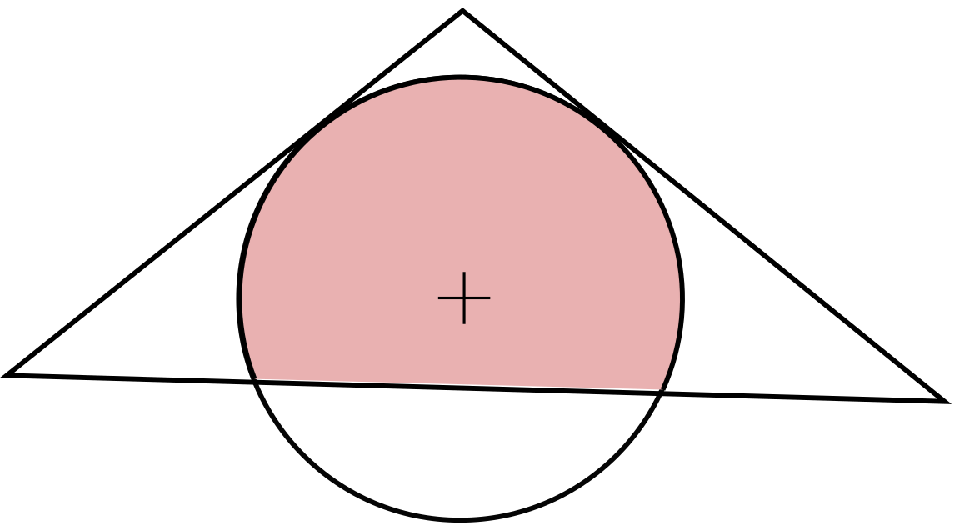}
&&
\includegraphics[scale=0.5]{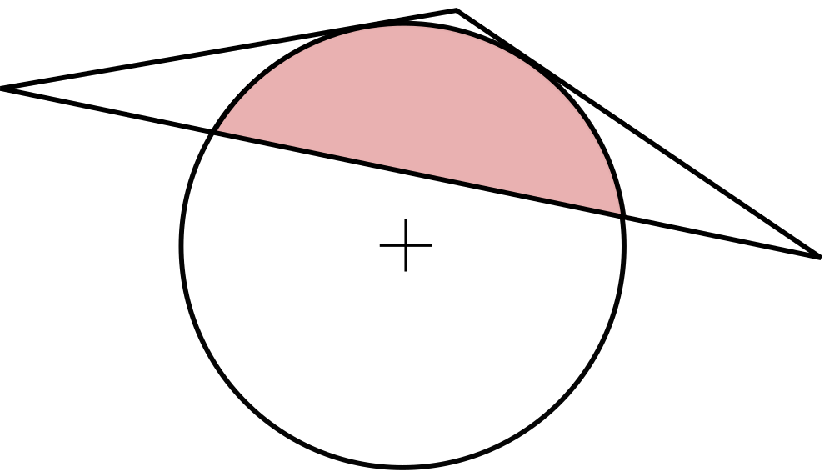}\\
(a)  && (b)\\
\includegraphics[scale=0.5]{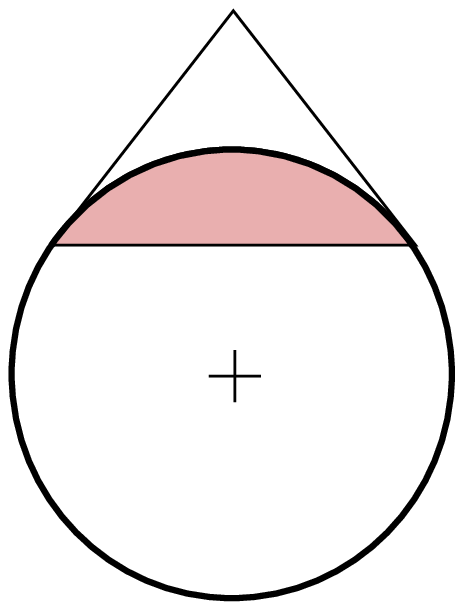}
&&
\includegraphics[scale=0.5]{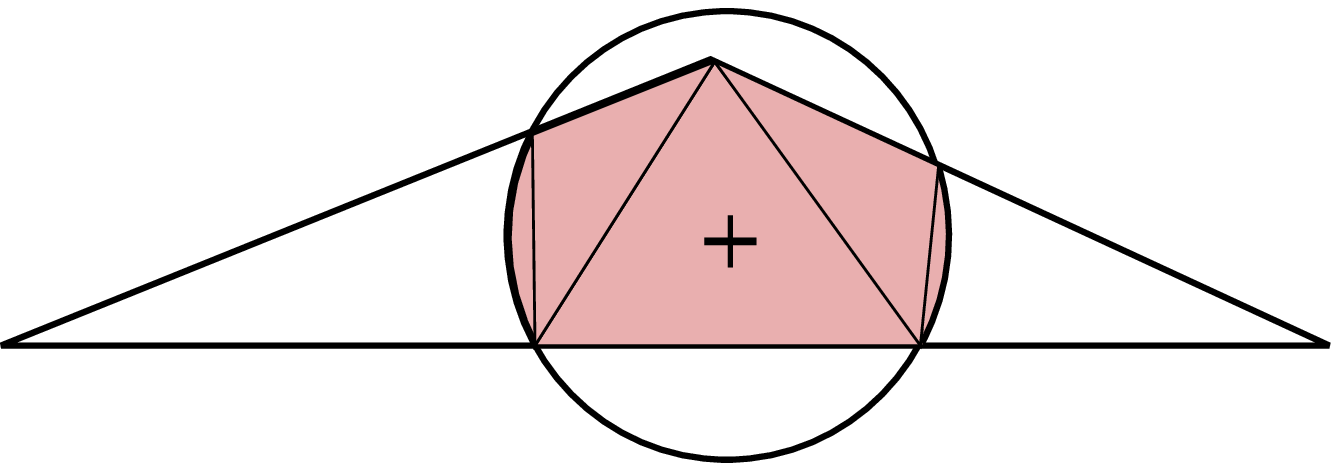}\\
(c)  && (d)\\
\includegraphics[scale=0.5]{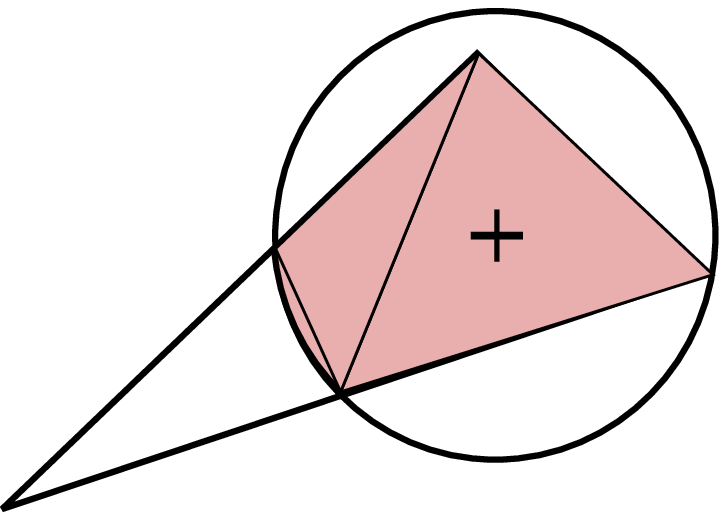}
&&

\includegraphics[scale=0.5]{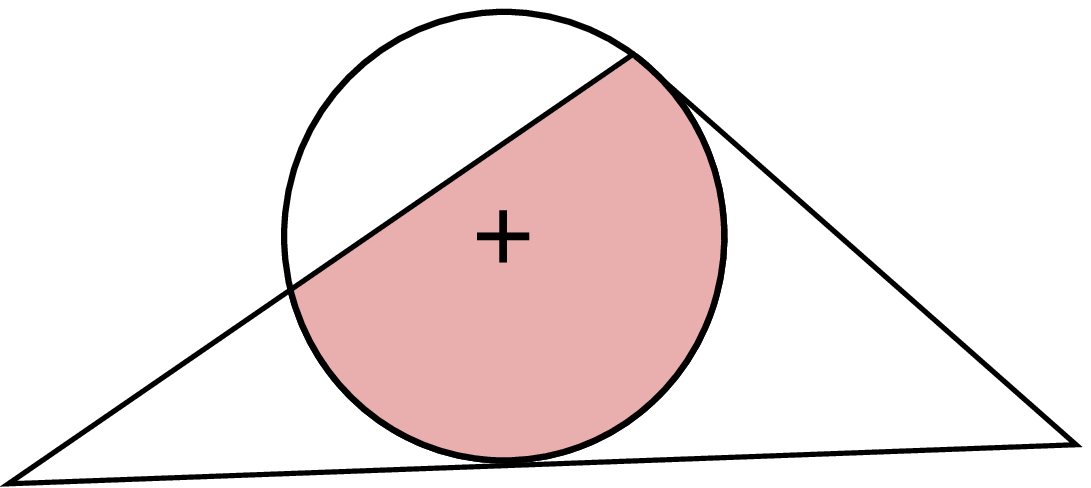}\\
(e)  && (f)\\
\includegraphics[scale=0.5]{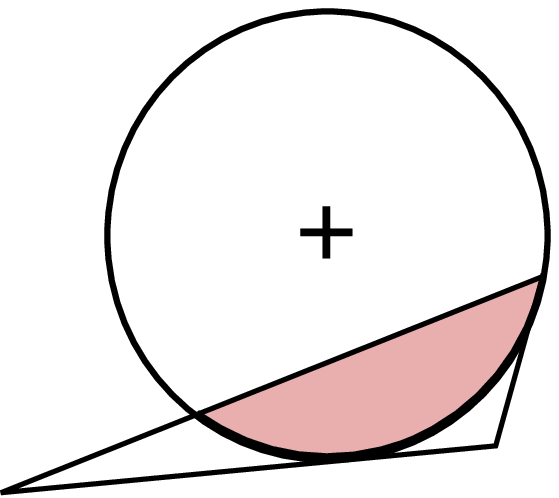}
&&\includegraphics[scale=0.5]{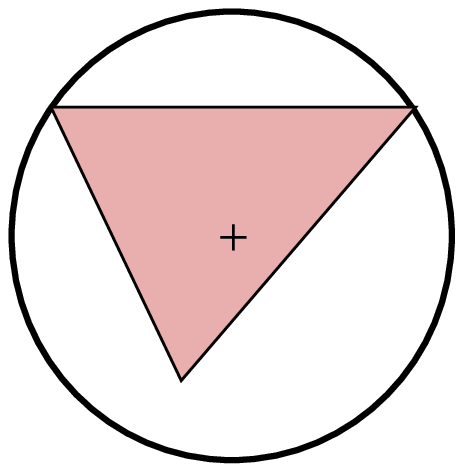}
\\
(g)&&(h)\\
\end{tabular}
\end{tabular}
\caption{
Possible shapes for the intersection when {\tt{Code=$14, 16$ or $22$}}.} 
\label{fig211}
\end{figure}

 $\bullet\; {\textbf{Code=14, 16, 22}}$
obtained when 
$(n_3, n_2, n_1)=(1,1,2), (1,2,1), (2,1,1)$.
The possible shapes for the intersection
are represented in Figure \ref{fig211}.
It follows that the pseudo-code for computing the area of the intersection
(stored in variable {\tt{area}}) when {\tt{Code}}$=14$
is the following:\\
\par {\tt{area}}$=0$.
\par {\textbf{if }}$d_A>R$ and $d_B>R$ and $d_C >R$ then 
\par \hspace*{0.4cm}{\tt{[out]}}={\tt{Are\_In}}\_{\tt{Same}}\_{\tt{Half}}\_{\tt{Space}}$(P,C,A_1,A_2)$,
\par \hspace*{0.4cm}{\textbf{if }}out=1 then 
{\tt{area}}$=\pi R^2 -L(P,R,A_1,A_2)$, 
\par \hspace*{0.4cm}{\textbf{else }}{\tt{area}}$=L(P,R,A_1,A_2)$,
\par \hspace*{0.4cm}{\textbf{end if}}
\par {\textbf{else if }}$d_A>R$ and $d_C<R$ and $d_B>R$ then
\par \hspace*{0.4cm}{\tt{area}}$=L(P,R,A_1,C_1)
+L(P,R,B_1,A_2)+T(C,C_1,A_1) 
+T(C,A_1,A_2)+T(C,B_1,A_2)$,
\par {\textbf{else if }}$d_B=R$ and 
$d_A=R$ and $d_C<R$ then {\tt{area}}$=T(A,B,C)$,
\par {\textbf{else if }}$d_B=R$ and 
$d_A=R$ and $d_C>R$ then {\tt{area}}$=L(P,R,A,B)$,

\par {\textbf{else if }}$d_B=R$
\par \hspace*{0.4cm}{\textbf{if }}$d_C<R$ then
{\tt{area}}$=L(P,R,A_1,C_1)+T(B,C,A_1)+T(C,A_1,C_1)$,
\par \hspace*{0.4cm}{\textbf{else }}
\par \hspace*{1.2cm}{\tt{[out]}}={\tt{Are\_In}}\_{\tt{Same}}\_{\tt{Half}}\_{\tt{Space}}$(P,C,B,A)$,
\par \hspace*{1.2cm}{\textbf{if }}{\tt{out}}$=1$ then {\tt{area}}$=\pi R^2 -L(P,R,A_1,B)$,
\par \hspace*{1.2cm}{\tt{else }}{\tt{area}}$=L(P,R,A_1,B)$,
\par \hspace*{1.2cm}{\textbf{end if}}
\par \hspace*{0.4cm}{\textbf{end if}}

\par {\textbf{else if }}$d_A=R$
\par \hspace*{0.4cm}{\textbf{if }}$d_C<R$ then
{\tt{area}}$=L(P,R,B_1,A_2)+T(A,C,A_2)+T(C,B_1,A_2)$,
\par \hspace*{0.4cm}{\textbf{else }}
\par \hspace*{1.2cm}{\tt{[out]}}={\tt{Are\_In}}\_{\tt{Same}}\_{\tt{Half}}\_{\tt{Space}}$(P,C,B,A)$,
\par \hspace*{1.2cm}{\textbf{if }}{\tt{out}}$=1$ then {\tt{area}}$=\pi R^2 -L(P,R,A,A_2)$,
\par \hspace*{1.2cm}{\tt{else }}{\tt{area}}$=L(P,R,A,A_2)$,
\par \hspace*{1.2cm}{\textbf{end if}}
\par \hspace*{0.4cm}{\textbf{end if}}
\par {\textbf{end if}}\\
The pseudo-codes when 
{\tt{Code}}$=16$ and $22$ are 
obtained
by appropriate permutation of $A,B,C$,
and the intersection points.\\
\par
\begin{figure}
\centering
\begin{tabular}{ccc}
\includegraphics[scale=0.5]{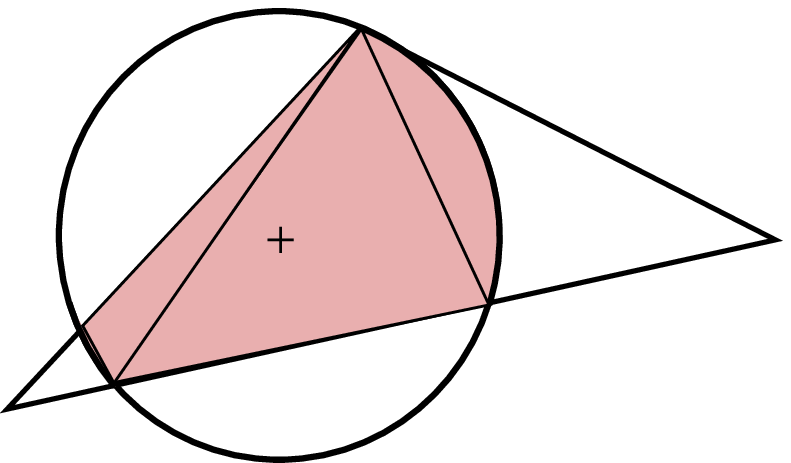}
&&
\includegraphics[scale=0.5]{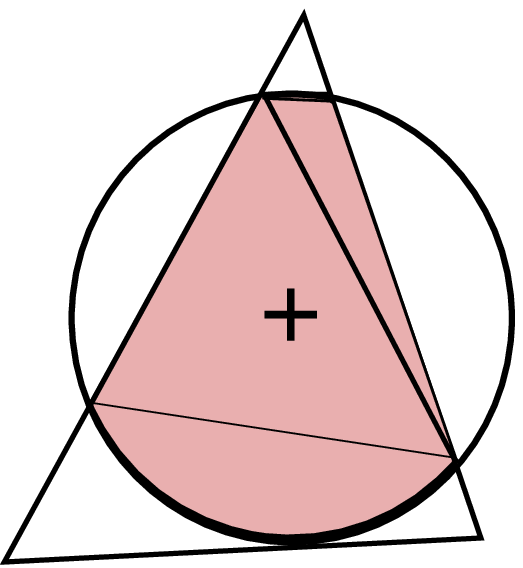}\\
(a)&  & (b) \\
\includegraphics[scale=0.5]{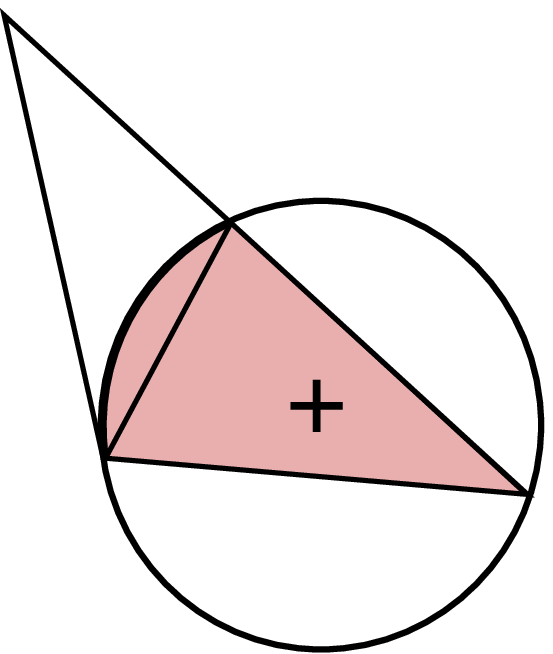}
&&
\includegraphics[scale=0.4]{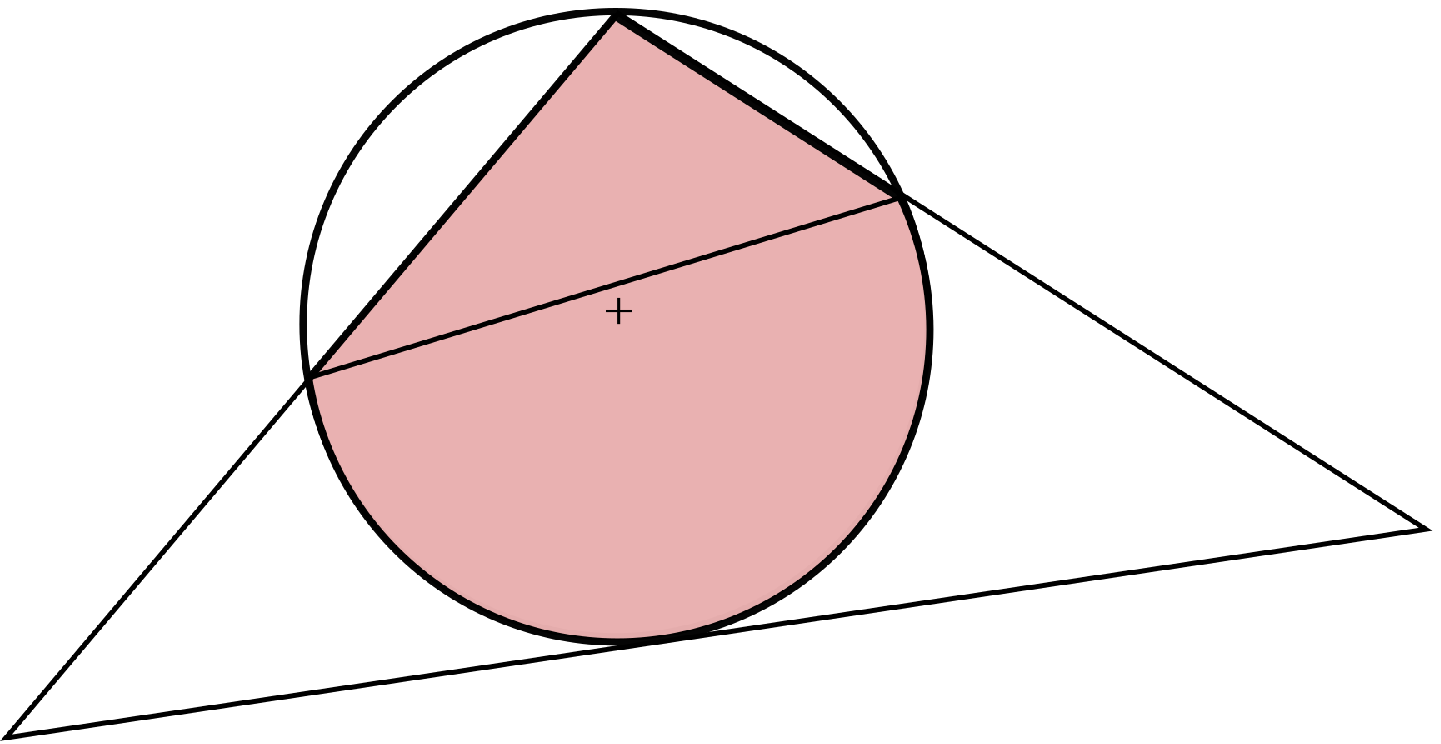}\\
(c)&  & (d) \\
\includegraphics[scale=0.4]{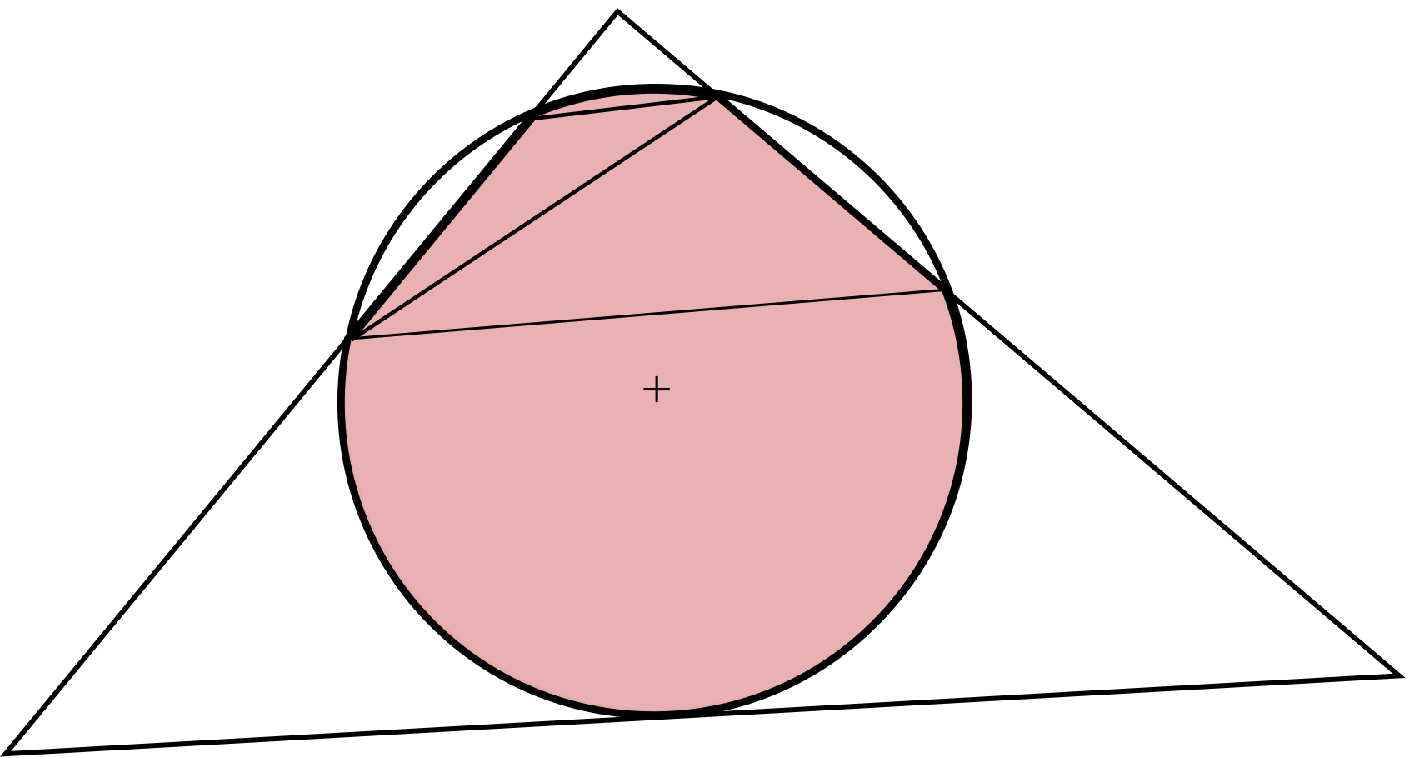}
&&\includegraphics[scale=0.4]{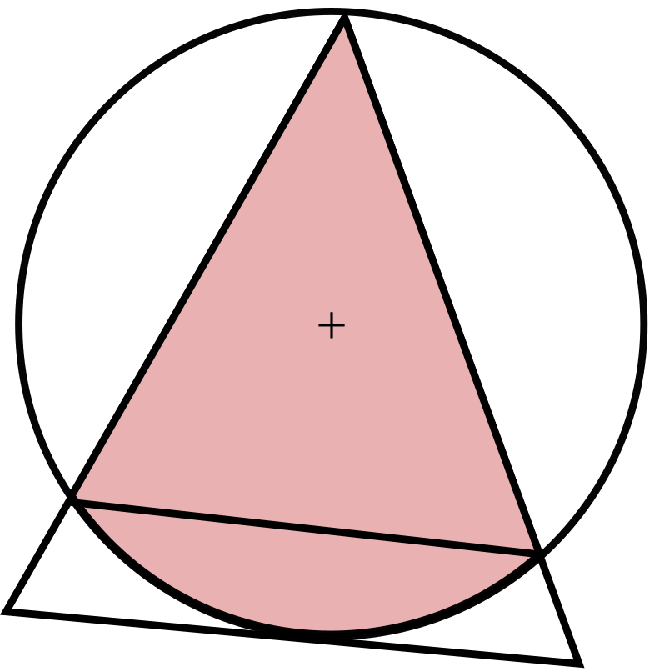}\\
(e) &&(f)\\
\end{tabular}
\caption{Possible shapes for the intersection when {\tt{Code=$17, 23$ or $25$}}.} 
\label{fig221}
\end{figure}

 $\bullet \; {\textbf{Code=17, 23, 25}}$
obtained when $(n_3,n_2,n_1)=(1,2,2),(2,1,2),(2,2,1)$.
The possible shapes for the intersection are represented
in Figure \ref{fig221}.
It follows that when {\tt{Code}}$=17$ the pseudo-code 
for computing the area of the intersection is the
following:\\

\par {\textbf{if }}$d_A>R$ and $d_B>R$ and $d_C>R$ then
\par \hspace*{0.4cm}{\tt{[out]}}={\tt{Are\_In}}\_{\tt{Same}}\_{\tt{Half}}\_{\tt{Space}}$(P,B,B_2,A_1)$,
\par \hspace*{0.4cm}{\textbf{if }}out$=1$ then
{\tt{area}}$=L(P,R,B_1,A_2)+L(P,R,A_1,B_2) + T(B_1,A_2,B_2)+T(A_1,A_2,B_2)$,
\par \hspace*{0.4cm}{\textbf{else }}
{\tt{area}}$=L(P,R,B_1,A_2)+\pi R^2 -L(P,R,A_1,B_2) + T(B_1,A_2,B_2)+T(A_1,A_2,B_2)$,
\par \hspace*{0.4cm}{\textbf{end if}}
\par {\textbf{else if }}$d_A=R$ and $d_B=R$ then {\tt{area}}$=L(P,R,A,B_2)+T(A,B,B_2)$,
\par {\textbf{else if }}$d_C=R$ and $d_B=R$ then {\tt{area}}$=L(P,R,C,A_1)+T(B,C,A_1)$,
\par {\textbf{else if }}$d_B=R$ then 
\par \hspace*{0.4cm}{\tt{[out]}}={\tt{Are\_In}}\_{\tt{Same}}\_{\tt{Half}}\_{\tt{Space}}$(P,B,B_2,A_1)$,
\par \hspace*{0.4cm}{\textbf{if }}{\tt{out}}$=1$ then
{\tt{area}}$=L(P,R,A_1,B_2) +T(A_1,B,B_2)$,
\par \hspace*{0.4cm}{\textbf{else }}
{\tt{area}}$=\pi R^2-L(P,R,A_1,B_2) +T(A_1,B,B_2) $,
\par \hspace*{0.4cm}{\textbf{end if}}
\par {\textbf{else if }}$d_A=R$ then
$$
{\tt{area}}=L(P,R,A_1,B_2)+L(P,R,A_2,B_1)+T(A,B_2,B_1)+T(A,B_1,A_2),
$$
\par {\textbf{else if }}$d_C=R$ then
$$
{\tt{area}}=L(P,R,A_1,C)+L(P,R,A_2,B_1)+T(C,A_2,A_1)+T(C,B_1,A_2),
$$
\par {\textbf{end if.}}\\
\par The pseudo-codes when 
{\tt{Code}}$=23$ and $25$ are 
obtained 
by appropriate permutation of $A,B,C$,
and the intersection points.\\

\par
\begin{figure}
\centering
\begin{tabular}{ccc}
\includegraphics[scale=0.5]{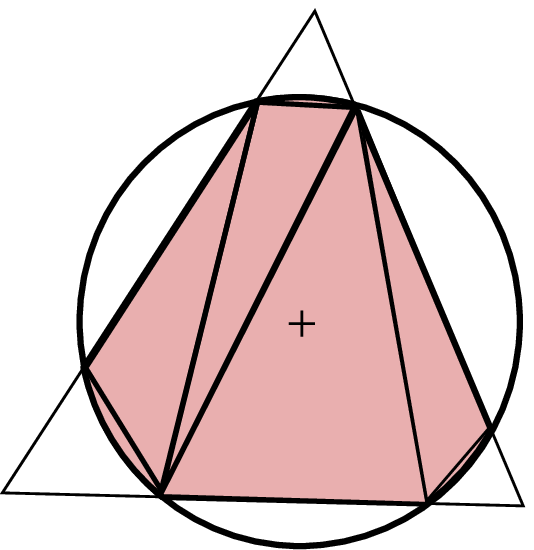}
&&
\includegraphics[scale=0.5]{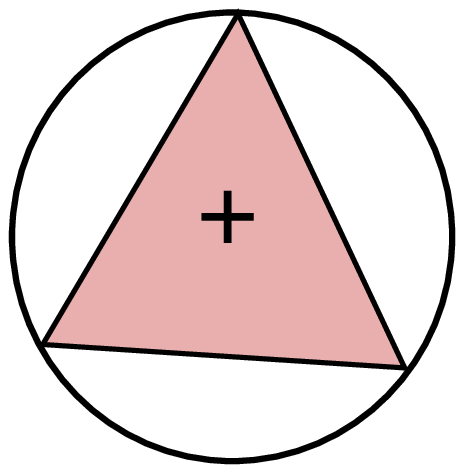}\\
(a)  &\hspace*{1cm} &(b)\\
\includegraphics[scale=0.5]{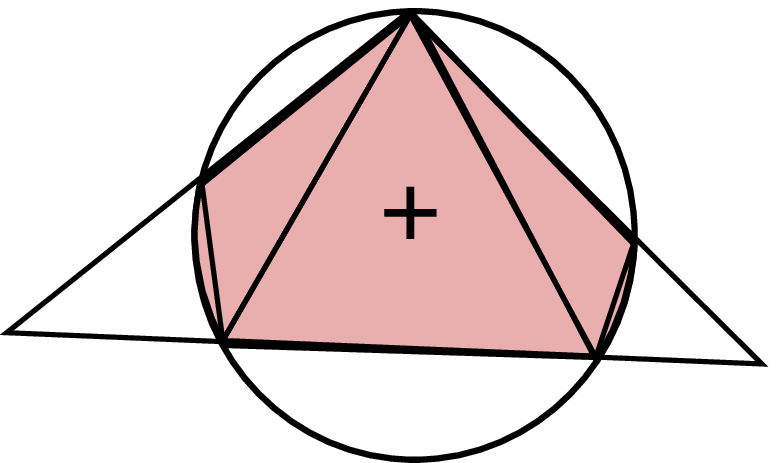}
&&
\includegraphics[scale=0.5]{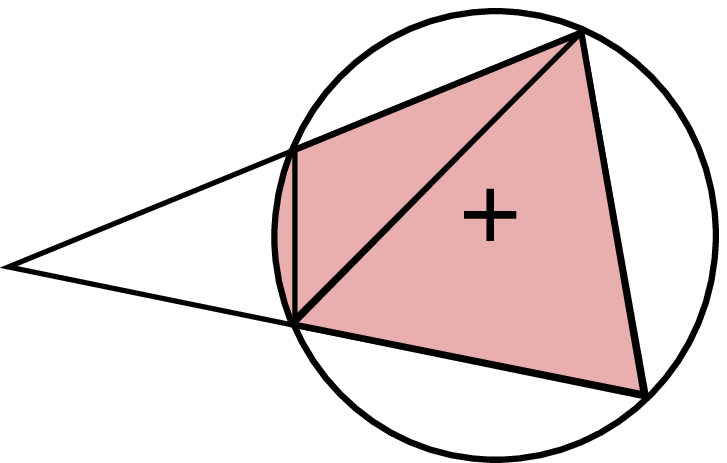}\\
(c)  & \hspace*{1cm} &(d) \\
\end{tabular}
\caption{
Possible shapes for the intersection when {\tt{Code=$26$}}.} 
\label{fig222}
\end{figure}

 $\bullet \;{\textbf{Code=26}}$
corresponding to 
$(n_3,n_2,n_1)=(2,2,2)$.
The possible shapes for the intersection
are represented in Figure \ref{fig222}.
From this figure we obtain the following pseudo-code
to compute the intersection area:
\par {\textbf{if }}$d_A>R$ and $d_B>R$ and $d_C>R$ then
$$
\begin{array}{lll}
{\tt{area}}&=&L(P,R,A_2,B_1)+L(P,R,B_2,C_1)+L(P,R,A_1,C_2)\\
&&+T(A_2,B_1,A_1)+T(A_1,C_2,B_1)+T(C_2,B_1,B_2)+T(B_2,C_1,C_2),
\end{array}
$$
\par {\textbf{else if}}$d_A=R$ and $d_B=R$ and $d_C=R$ then {\tt{area}}$=T(A,B,C)$,
\par {\textbf{else if }}$d_A=R$ and $d_B=R$ and $d_C>R$ then
$$
{\tt{area}}=L(P,R,C_1,B_2)+T(A,B,C_1)+T(B,C_1,B_2),
$$
\par {\textbf{else if }}$d_A=R$ and $d_C=R$ and $d_B>R$ then
$$
{\tt{area}}=L(P,R,A_2,B_1)+T(A,C,A_2)+T(C,A_2,B_1),
$$
\par {\textbf{else if }}$d_B=R$ and $d_C=R$ and $d_A>R$ then
$$
{\tt{area}}=L(P,R,A_1,C_2)+T(C,B,A_1)+T(C,A_1,C_2),
$$
\par {\textbf{else if }}$d_A=R$ and $d_B>R$ and $d_C>R$ then
$$
{\tt{area}}=L(P,R,A_2,B_1)+L(P,R,B_2,C_1)+T(A,A_2,B_1)+T(A,B_1,B_2)+T(A,C_1,B_2),
$$
\par {\textbf{else if }}$d_B=R$ and $d_A>R$ and $d_C>R$ then
$$
{\tt{area}}=L(P,R,A_1,C_2)+L(P,R,C_1,B_2)+T(A,B,C_2)+T(B,C_1,C_2)+T(B,B_2,C_1),
$$
\par {\textbf{else if }}$d_C=R$ and $d_A>R$ and $d_B>R$ then
$$
{\tt{area}}=L(P,R,A_1,C_2)+L(P,R,A_2,B_1)+T(C,C_2,A_1)+T(C,A_1,A_2)+T(C,B_1,A_2),
$$
\par {\textbf{end if}}

\section{The library: main functions and examples}\label{lib}

The Matlab library, available at \url{https://github.com/vguigues/Areas_Library}, computes the density 
of the distance $D$ to a random
variable uniformly distributed in some sets and the area
of intersection of disks and balls with those sets.

All necessary files to run the functions of the library
are in folders {\tt{Areas\_Library}} and its subfolder
{\tt{Examples}} which contains files to run the main functions
on examples. No external library is needed.
We implemented all functions of these folders 
except function {\tt{polygon\_triangulate}} 
which computes a triangulation of a polygone.
This function, which can be found in folder {\tt{Areas\_Library}},
is the Matlab version by John Burkardt of the original
C version by Joseph ORourke \cite{rourkebook}.

If the folder {\tt{Areas\_Libary}} is copied in folder
{\tt{C:\char`\\Users\char`\\user\_name}}, before using the library, update the path  in Matlab with commands:\\

\par {\tt{addpath 'C:\char`\\Users\char`\\user\_name\char`\\Areas\_Libary'}}
\par {\tt{addpath 'C:\char`\\Users\char`\\user\_name\char`\\Areas\_Libary\char`\\Examples'}}\\

\par The next section shows how to use the main functions of the library
using the files of examples that can be found in folder
{\tt{Areas\_Libary\char`\\Examples}}.

\subsection{Density of the distance
to a random variable uniformly distributed in a polygone}

The function to compute the density of the distance from a point $P \in \mathbb{R}^2$
to a random variable uniformly distributed in a polygone $S$ is:\\
\par {\tt{[d,time,dmin,dmax]=density\_polyhedron(S,P,Np,algo)}}\\
\par where input variables are:
\begin{itemize}
\item {\tt{algo}}: a char indicating the algorithm used. It can
take two values 'g' and 't1'.
When algo='g', the algorithm given in 
\cite{guiguesarxivcg2015} based on Green's theorem is used.
When algo='t1' the algorithm from Section \ref{algotriang1} is used
to compute the intersection areas of disks and the polygone.
\item {\tt{Np}}: the number of discretization points: the density is computed at
{\tt{Np}} equally spaced points $x_1,x_2,\ldots,x_{\mbox{\tt{Np}}}$ from the support of the random variable distance.
\item $P$: point $P$ as explained above.
\item $S=[S_1;S_2;S_3;\ldots;S_n;S_1]$: the polyhedron
where $n$ is the number of vertices and $S_1, S_2, S_3,\ldots,S_n$
are the successive vertices of the polyhedron. Observe that the first point $S_1$ is repeated.
When algo='g', when travelling on the boundary of $S$ from $S_1$ to $S_2$, 
then from $S_2$ to $S_3$ and so on until the last line segment $\overline{S_n S_1}$, one always has 
the relative interior of $S$ to the left.
When algo='t1' this restriction does not apply: if algo='t1', 
when travelling on
of $S$ from $S_1$ to $S_2$, 
then from $S_2$ to $S_3$ and so on until the last line segment $\overline{S_n S_1}$, one 
can either  have
the relative interior of $S$ to the left or to the right.\\
\end{itemize}
Output variables of function  {\tt{density\_polyhedron}} are:
\begin{itemize}
\item {\tt{d}}: a vector of size {\tt{Np}} where $d(i)$ is the estimation of the density
of the random variable at $x_i$.
\item {\tt{time}} is the time required to compute $d$.
\item ]{\tt{dmin}},{\tt{dmax}}[ is the support of the random variable meaning that 
{\tt{dmax}} is the maximal distance between $P$ and the boundary of the polygone.
If $P$ is inside the polygone then 
{\tt{dmin}}$=0$ and if $P$ is outside the polygone
then {\tt{dmin}} is the minimal distance from $P$ to the boundary of the polygone.
\end{itemize}

We illustrate the use of this function on several examples 
written
in folder {\tt{'Areas\_Libary\char`\\Examples'}}.

We start with an example written in file {\tt{drectex.m}}
of folder {\tt{Examples}} where $S$ is a rectangle with side lengths
$L$ and $\alpha L$ with $0<\alpha<1$
and $P$ is the center of the rectangle.  
For this example, the density of the distance from $P$ to a random variable
uniformly distributed in $S$ is known in closed form and is given in \cite{stewartzhang2012}.
Therefore, this example allows us to test the implementation
of function {\tt{density\_polyhedron}} comparing output {\tt{d}} of this function
when algo='g', 't1' with the theoretical values given in \cite{stewartzhang2012}.

The function corresponding to this example is\\

\par {\tt{[dG,dT1,d,ErrG,ErrT1]=drectex(Np,$\alpha$,$L$)}}.\\

The input variables are parameters {\tt{Np}}, $\alpha, L$, given above and the outputs
are the following:
\begin{itemize}
\item {\tt{dG}} and {\tt{dT1}} are vectors of size {\tt{Np}} and 
{\tt{dG}}$(i)$ (resp. {\tt{dT1}}$(i)$) is the value of the density at $x_i,i=1,\ldots,${\tt{Np}}, computed
calling function 
{\tt{density\_polyhedron}} with variable algo='g'
(resp. calling function 
{\tt{density\_polyhedron}} with variable algo='t1').
Recall that $x_i,i=1,\ldots,x_{\mbox{\tt{Np}}}$ are {\tt{Np}} equally spaced points
in ]{\tt{dmin}},{\tt{dmax}}[.
\item {\tt{d}} is a vector of size 
{\tt{Np}}: {\tt{d}}$(i)$ is the exact value of the density at $x_i$ computed
using the analytic formulas given in \cite{stewartzhang2012}.
\item {\tt{ErrG}} is the maximal error when algo='g', i.e.,
{\tt{ErrG}}$=\max_{i=1,\ldots,\mbox{\tt{Np}}} |\mbox{\tt{dG}}(i)-\mbox{\tt{d}}( i )|$.
\item {\tt{ErrT1}} is the maximal error when algo='t1', i.e.,
{\tt{ErrT1}}$=\max_{i=1,\ldots,\mbox{{\tt{Np}}}} |\mbox{\tt{dT1}}(i)-\mbox{\tt{d}}( i )|$.
\end{itemize}
On top of that, the function plots
vectors {\tt{dG}}, {\tt{dT1}}, and {\tt{d}}.
For instance, running\\
\par
\begin{figure}
\centering
\includegraphics[scale=1.2]{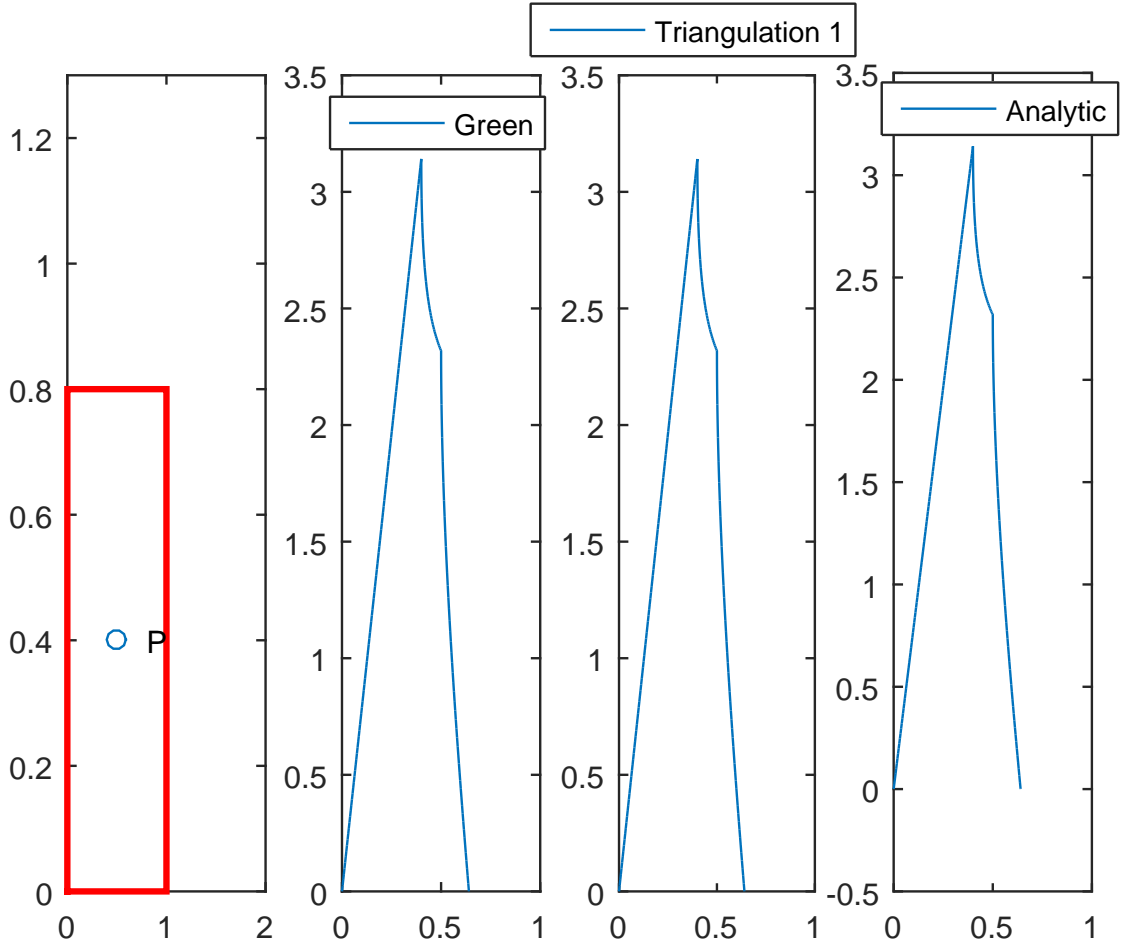}
\caption{Plots produced calling 
{\tt{[dG,dT1,d,ErrG,ErrT1]=drectex(Np,$\alpha$,$L$)}}.
From left to right: rectangle $S$ and $P$,
plot of dG, plot of dT1, and plot of d.} 
\label{ex2legend}
\end{figure}

 {\tt{[dG,dT1,d,ErrG,ErrT1]=drectex(1,0.8,1)}}.\\
\par the plots of Figure  \ref{ex2legend} are displayed.
On this Figure, from left to right, the first plot represents
rectangle $S=[(0,0);(1,0);(1,0.8);(0,0.8);(0,0)]$ and $P=(0.5,0.4)$,
the second plot represents the density of $D$ obtained using the algorithm
from \cite{guiguesarxivcg2015}, the third plot is the density of $D$
obtained using the algorithm from Section \ref{algotriang1},
while the last plot is the graph of the true density of $D$.

In this example, the densities are computed with the following Matlab code:\\

\par {\tt{P1=[0,0]; P2=[L,0]; P3=[L,alpha$\times$L]; P4=[0,alpha$\times$L];}}
\par {\tt{P=[L/2,alpha$\times$L/2];}}
\par {\tt{S=[P1;P2;P3;P4;P1];}}
\par {\tt{[dG,timeg,dmin,dmax]=density\_polyhedron(S,P,Nb,'g')}};
\par {\tt{[dT1,timet1,dmin,dmax]=density\_polyhedron(S,P,Nb,'t1')}};
\par {\tt{ErrG=max(abs(d-dG)); ErrT1=max(abs(d-dT1)).}}\\

To check the implementations of the algorithm from \cite{guiguesarxivcg2015}
and the algorithm from Section \ref{algotriang1}, we now report 
in Table \ref{tablenum1}
the values of {\tt{ErrG}} and
{\tt{ErrT1}} for several values of 
the number {\tt{Np}} of discretization points, namely when
{\tt{Np}} varies in the set $\{10\,000, 20\,000, 50\,000, 100\,000\}$.


\begin{table}
\begin{tabular}{|c|c|c|}
\hline
{\tt{Np}}   & {\tt{ErrG}} & {\tt{ErrT1}} \\
\hline
$10\,000$ &  0.017  &  0.023  \\ 
\hline 
$20\,000$ &  0.010  &  0.020  \\
\hline
$50\,000$ &  0.007  & 0.04   \\
\hline   
$100\,000$ & 0.004   & 0.03   \\
\hline   
\end{tabular}
\caption{Maximal error obtained with the algorithms from \cite{guiguesarxivcg2015} and Section \ref{algotriang1} to
 compute the density of
$D$ ($D$ being the distance from the center of a rectangle
with side lengths 1 and 0.8 to a random variable with uniform
distribution in this rectangle) at {\tt{Np}} discretization points.}\label{tablenum1}
\end{table}
In all cases the maximal error is very small which shows that 
both algorithms  correctly compute 
the {\tt{Np}} areas of  intersection of the disks and polygone of this 
example.\footnote{To approximate the density at 
{\tt{Np}} points, we need to compute the cumulative distribution function 
at {\tt{Np}} points and therefore when 
{\tt{Np}}$=100\,000$, the algorithms  
are called $100\,000$ times each to compute $100\,000$ areas.}
We also observe that the approximations are slightly better with
the algorithm from \cite{guiguesarxivcg2015} and, as expected, the maximal error decreases with 
{\tt{Np}} for this algorithm. This is not the case for the other, probably
due to roundoff errors.\\

\par We now compare the algorithms 
on  other examples coded in Matlab file {\tt{dpolyex.m}}.
 More precisely, we consider three polyhedra
(a triangle, a rectangle, and an arbitrary polygone) 
and in each case a point $P$
inside the polygone and a point $P$ outside.
For these 6 examples the Matlab codes are the following.\\
\par
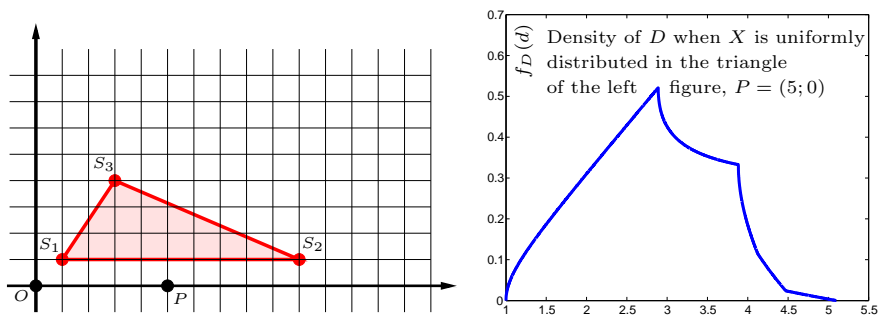
\begin{figure}[H]
\begin{tabular}{ll}
\input{Triangle.pstex_t} &  \input{Triangle_5_0B.pstex_t}
\end{tabular}
\caption{Density of $D$ when $X$ is uniformly distributed in a triangle.} 
\label{figuredenspolyhedron1}
\end{figure}

 $\bullet$ $S=[( 1,1);(10,1);(3,4);(1,1)]$ is a triangle and $P=[5;0]$ is outside this
triangle. In Figure \ref{figuredenspolyhedron1},
$S$ and $P$ are represented in 
the left plot while
the corresponding density of $D$ is represented
in the right plot.
This density is obtained with the following Matlab code:\\
\par {\tt{P1=[1,1]; P2=[10,1]; P3=[3,4]; S=[P1;P2;P3;P1]; P=[5,0];}}
\par {\tt{[dT1,timeT1,dminT1,dmaxT1]=density\_polyhedron(S,P,Np,algo);}}\\
\par where algo='g' or 't1'.\\
\par
\begin{figure}[H]
\begin{tabular}{ll}
\input{TriangleIn.pstex_t} &  \input{Triangle_4_2B.pstex_t}
\end{tabular}
\caption{Density of $D$ when $X$ is uniformly distributed in a triangle.} 
\label{figuredenspolyhedron2}
\end{figure}
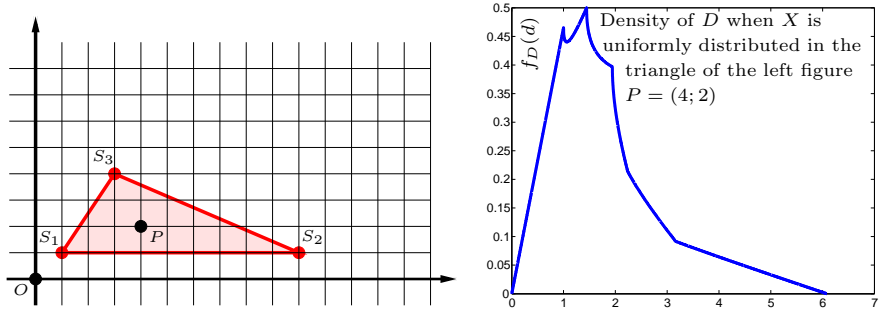

 $\bullet$ $S=[( 1,1);(10,1);(3,4);(1,1)]$ is a triangle and $P=[4;2]$ is inside this
triangle. In Figure \ref{figuredenspolyhedron2},
$S$ and $P$ are represented in 
the left plot while
the corresponding density of $D$ is represented
in the right plot.
This density is obtained with the following Matlab code:\\

\par {\tt{P1=[1,1]; P2=[10,1]; P3=[3,4]; S=[P1;P2;P3;P1];}}; {\tt{P=[4,2];}}
\par {\tt{[dT2,timeT2,dminT2,dmaxT2]=density\_polyhedron(S,P,Np,algo);}}\\

\par where algo='g' or 't1'.\\
\par
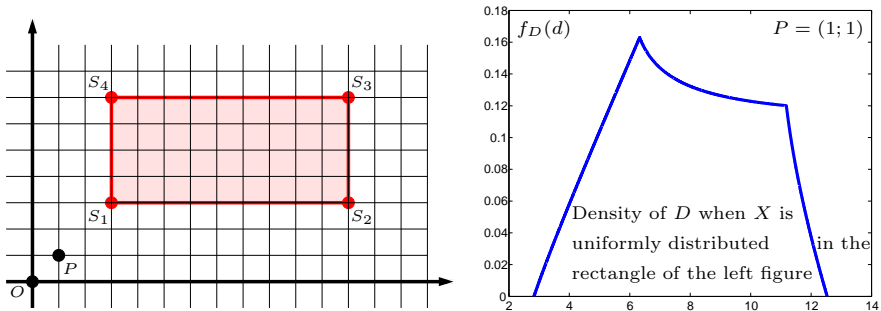
\begin{figure}[H]
\begin{tabular}{ll}
\input{Rectangle.pstex_t} &  \input{Rectangle_1_1B.pstex_t}
\end{tabular}
\caption{Density of $D$ when $X$ is uniformly distributed in a rectangle.} 
\label{figuredenspolyhedron3}
\end{figure}

 $\bullet$ $S=[(3,3);(12,3);(12,7);(3,7);(3,3)]$ is a rectangle and 
$P=[1;1]$ is outside this
rectangle. In Figure \ref{figuredenspolyhedron3},
$S$ and $P$ are represented in 
the left plot while
the corresponding density of $D$ is represented
in the right plot.
This density is obtained with the following Matlab code:\\

\par {\tt{P1=[3,3]; P2=[12,3]; P3=[12,7]; P4=[3,7]; S=[P1;P2;P3;P4;P1]; P=[1,1];}}
\par {\tt{[dR1,timeR1,dminR1,dmaxR1]=density\_polyhedron(S,P,Np,algo);}}\\
\par where algo='g' or 't1'.\\

\par
\begin{figure}[H]
\begin{tabular}{ll}
 \input{RectangleIn.pstex_t} &  \input{Rectangle_6_5B.pstex_t}
\end{tabular}
\caption{Density of $D$ when $X$ is uniformly distributed in a rectangle.} 
\label{figuredenspolyhedron4}
\end{figure}
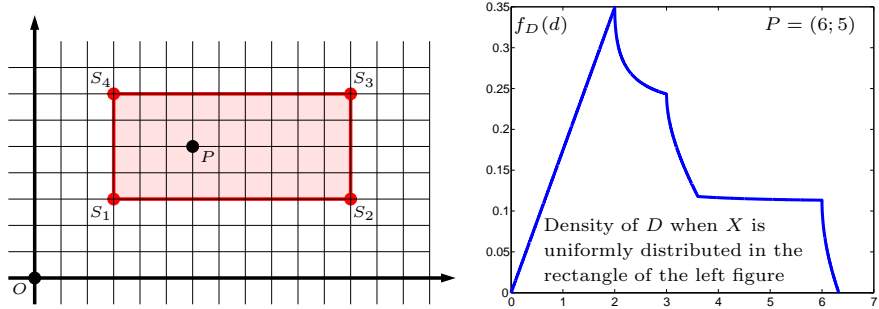

 $\bullet$ $S=[(3,3);(12,3);(12,7);(3,7);(3,3)]$ is a rectangle and 
$P=[6;5]$ is inside this
rectangle. In Figure \ref{figuredenspolyhedron4},
$S$ and $P$ are represented in 
the left plot while
the corresponding density of $D$ is represented
in the right plot.
This density is obtained with the following Matlab code:\\

\par {\tt{P1=[3,3]; P2=[12,3]; P3=[12,7]; P4=[3,7]; S=[P1;P2;P3;P4;P1]; P=[6,5];}}
\par {\tt{[dR2,timeR2,dminR2,dmaxR2]=density\_polyhedron(S,P,Np,algo);}}\\
\par where algo='g' or 't1'.\\

\par
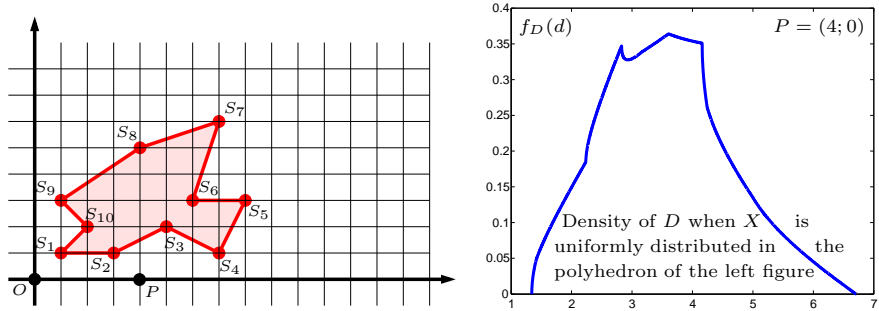
\begin{figure}[H]
\begin{tabular}{ll}
 \input{Poly_Ext.pstex_t} &  \input{Poly_40B.pstex_t}\\ 
\end{tabular}
\caption{Density of $D$ when $X$ is uniformly distributed in a polygone.} 
\label{figuredenspolyhedron5}
\end{figure}

 $\bullet$ $S=[(1,1);(3,1); (5,2); (7,1);(8,3);(6,3);(7,6);(4,5);(1,3);(2,2);(1,1)]$ is a polygone and 
$P=[4;0]$ is outside this
polygone. In Figure \ref{figuredenspolyhedron5},
$S$ and $P$ are represented in 
the left plot while
the corresponding density of $D$ is represented
in the right plot.
This density is obtained with the following Matlab code:\\
\par {\tt{P1=[1,1]; P2=[3,1]; P3=[5,2]; P4=[7,1]; P5=[8,3];}}
\par {\tt{P6=[6,3]; P7=[7,6]; P8=[4,5]; P9=[1,3]; P10=[2,2];}}
\par {\tt{S=[P1;P2;P3;P4;P5;P6;P7;P8;P9;P10;P1];}}
\par {\tt{P=[4,0];}}
\par {\tt{[dP1,timeP1,dminP1,dmaxP1]=density\_polyhedron(S,P,Np,algo);}}\\
\par where algo='g' or 't1'.\\

\par
\begin{figure}[H]
\begin{tabular}{ll}
 \input{Poly_Int.pstex_t} &  \input{Poly_43B.pstex_t}
\end{tabular}
\caption{Density of $D$ when $X$ is uniformly distributed in a polygone.} 
\label{figuredenspolyhedron6}
\end{figure}
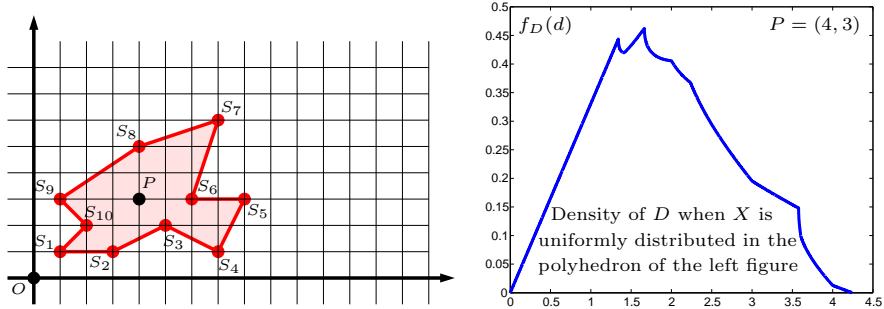

 $\bullet$ $S=[(1,1);(3,1);(5,2);(7,1);(8,3);(6,3);(7,6);(4,5);(1,3);(2,2);(1,1)]$ is a polygone and 
$P=[4;3]$ is inside this
polygone. In Figure \ref{figuredenspolyhedron6},
$S$ and $P$ are represented in 
the left plot while
the corresponding density of $D$ is represented
in the right plot.
This density is obtained with the following Matlab code:\\
\par {\tt{P1=[1,1]; P2=[3,1]; P3=[5,2]; P4=[7,1]; P5=[8,3];}}
\par {\tt{P6=[6,3]; P7=[7,6]; P8=[4,5]; P9=[1,3]; P10=[2,2];}}
\par {\tt{S=[P1;P2;P3;P4;P5;P6;P7;P8;P9;P10;P1];}}
\par {\tt{P=[4,3];}}
\par {\tt{[dP2,timeP2,dminP2,dmaxP2]=density\_polyhedron(S,P,Np,algo);}}\\
\par where algo='g' or 't1'.\\

Command\\
\par
\begin{figure}
\centering
\includegraphics[scale=1.25]{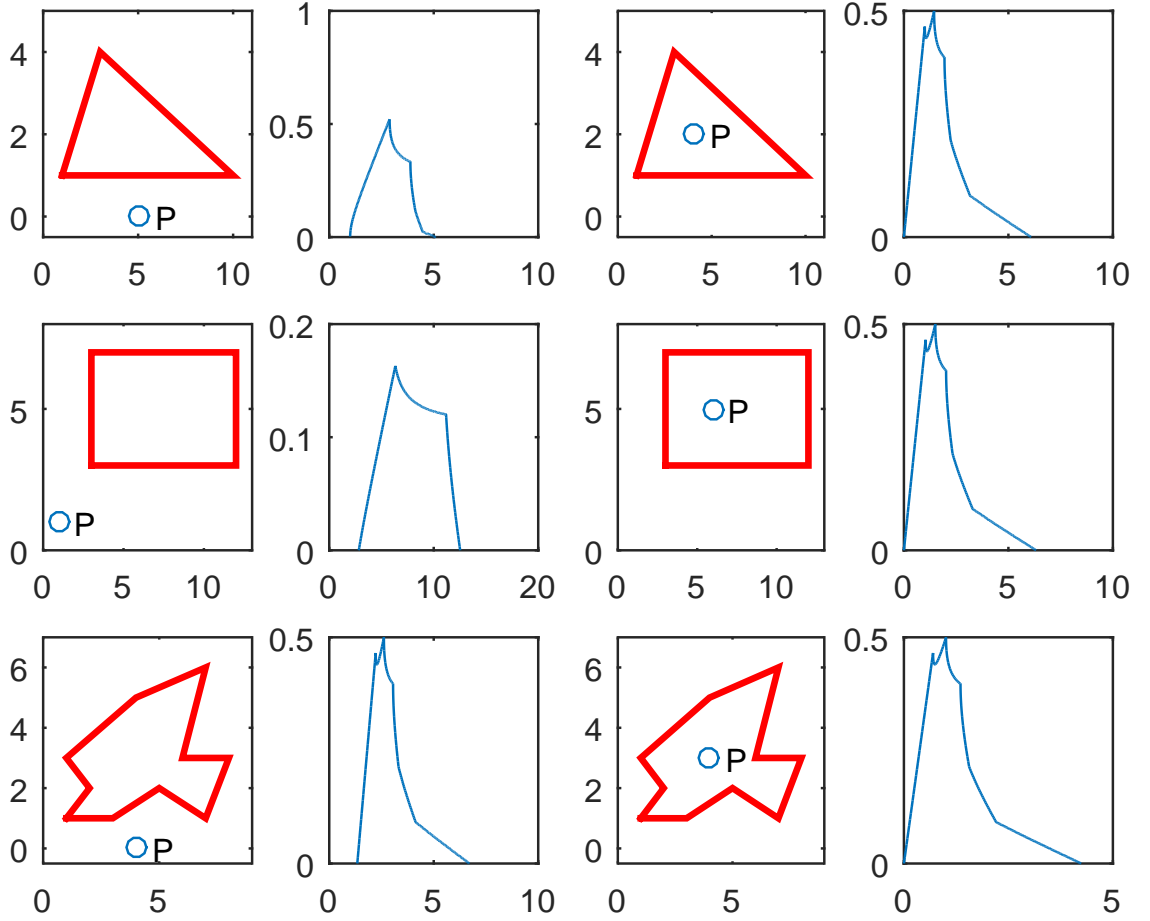}
\caption{Plots produced calling 
{\tt{[dT1,dT2,dR1,dR2,dP1,dP2]=dpolyex(10\,000,'g')}}.
In red, polygones $S$ with the corresponding densities of $D$
on their right (plots of dT1, dT2 on top, dR1, dR2 in the middle,
and dP1, dP2 at the bottom).} 
\label{ex1legend}
\end{figure}

 {\tt{[dT1,dT2,dR1,dR2,dP1,dP2]=dpolyex(10\,000,'g')}}\\
\par will run the code above to compute {\tt{dT1,dT2,dR1,dR2,dP1,dP2}}
with algo='g', {\tt{Np}}$=10\,000$, and will
produce Figure \ref{ex1legend} which represents polygones $S$
above and the corresponding densities of $D$ on their right.\\
\par Command\\
\par {\tt{[dT1,dT2,dR1,dR2,dP1,dP2]=dpolyex(Np,'t1')}}\\
\par does the same with algo='t1'.\\

\par Let $f_G( x_ i)$ (resp. $f_T( x_i )$) be the approximation of the density computed by 
the algorithm from \cite{guiguesarxivcg2015} based on Green's theorem (resp. the algorithm
from Section \ref{algotriang1}, based on a triangulation of the polygone) at $x_i$.
The maximal errors $\max_{i=1,\ldots,\mbox{{\tt{Np}}}} |f_G(x_i) - f_T(x_i)|$
were 
$5.7\small{\times}10^{-10}$ for $S, P$ given in Figure 
\ref{figuredenspolyhedron1}, $4.1\small{\times}10^{-8}$
for $S, P$ given in Figure 
\ref{figuredenspolyhedron2},
$8.6\small{\times}10^{-10}$ for $S, P$ given in Figure 
\ref{figuredenspolyhedron3},
$4.5\small{\times}10^{-10}$ for $S, P$ given in Figure 
\ref{figuredenspolyhedron4},
$2.3\small{\times}10^{-5}$ for $S, P$ given in Figure 
\ref{figuredenspolyhedron5}, and
$1.7\small{\times}10^{-9}$ for $S, P$ given in Figure 
\ref{figuredenspolyhedron6}.

The fact that these errors are very small is an indication that
both algorithms were correctly implemented.

\subsection{Area of the intersection of a disk and a polygone}

The area of the intersection of polygone 
$$
S=[S_1;S_2;\ldots;S_n;S_1]$$ (in Matlab notation)
and the disk of center $P \in \mathbb{R}^2$ and radius $d$ is computed as follows
with the library:\\
\par {\tt{[Crossing\_Number,AreaP,dmin,dmax]=polyhedron(S,P,n)}}
\par {\tt{[area]=area\_intersection\_disk\_polygone(S,P,d,n,Crossing\_Number,AreaP,algo)}}\\
\par where output {\tt{area}}
of function
{\tt{area\_intersection\_disk\_polygone}}
is the area of the intersection
and
the outputs of the first function
{\tt{polyhedron}} are:
\begin{itemize}
\item {\tt{Crossing\_Number}}: the crossing number for $S$ and $P$;
\item {\tt{AreaP}}: the area of polygone $S$;
\item {\tt{dmin}} (resp. {\tt{dmax}}): the minimal (resp. maximal) distance from $P$
to the border of the polygone.
\end{itemize}
When algo='g' (resp. 't1') the area of the intersection is computed
with the algorithm described in \cite{guiguesarxivcg2015} (resp. the algorithm
given in Section \ref{algotriang1}).

We test this function computing the
areas of intersection of 350 disks and polygones as well 
as the mean and maximal time required to compute these areas.
The polygones are generated using function\\
\par {\tt{[Polygone]=generate\_polygone(n,$R_0$)}}\\
\par of the library where parameters $n$ and $R_0$ are
described below.
This function generates randomly a polygone 
with $4n$ vertices
as follows.
We sample $4 n$ points 
taking $n$ points in each orthant with polar angles generated
randomly and independently in this orthant and
radial coordinates generated randomly and independently in 
the interval [0,$R_0$] (we take $R_0=1000$ in our experiments).
We then sort in ascending order the polar angles of these points.
This list defines the successive vertices of a 
star-shaped (simple) polygone.
The coordinates of the centers of the disks (resp. the radii) 
are obtained sampling
independently from the uniform distribution on the interval
$[-100,100]$ (resp. $[50,250]$). 

For each value of $n$ in the set $\{10, 25, 50, 80, 100, 150, 200\}$
we generate 50 star-shaped polyhedra and disks as explained above and for each polygone
and disk, we compute the area of their intersection using 
both algorithms.
The corresponding function of the library is\\
\par {\tt{[Errmax,ErrMoy,TimeGreen,TimeTr1]=random\_areas($M$)}}\\
\par where $M$ is the number of Monte-Carlo simulations ($M=50$ in our
experiments) and where the outputs are the following:
\begin{itemize}
\item {\tt{TimeGreen(k,j)}} 
(resp. {\tt{TimeTr1(k,j}}) is the time required
to compute the intersection area for $k$-th instance and
$j$-th value of $n$ (for instance for $j=1$ we have $n=10$, for
$j=2$, we have $n=25$) when
{\tt{area\_intersection\_disk\_polygone}} is called with
 algo='g' (resp. algo='t1');
\item {\tt{Errmax}} and {\tt{ErrMoy}}
are vectors of size $7$.
{\tt{ErrMoy}}$(j)$ and {\tt{Errmax}}$(j)$
are
defined respectively
by $\frac{1}{50}\sum_{k=1}^{50} |\mathcal{A}_G(k,j) - \mathcal{A}_T(k,j)|$  and  
$\max_{k=1,\ldots,50} |\mathcal{A}_G(k,j)-\mathcal{A}_T(k,j)|$ where 
$\mathcal{A}_G(k,j)$ and  $\mathcal{A}_T(k,j)$ are the
areas  of the intersection for 
$k$-th Monte-Carlo simulation and $j$-th value of $n$
computed with respectively algo='g' and algo='t1'. 
\end{itemize}
For each value of $n$, the mean and maximal time (over the 50 instances)
required to compute these areas are reported in Table \ref{tablerandompoly}.
We also report in this table  the values of 
{\tt{Errmax}} and {\tt{ErrMoy}}.
\begin{table}
\begin{tabular}{|c|c|c|c|c|c|c|}
\hline
$4n$  &  \begin{tabular}{c}Mean time\\algo='t1'\end{tabular} &\begin{tabular}{c} Mean Time\\algo='g' \end{tabular}& 
\begin{tabular}{c}Max time\\algo='t1'\end{tabular}&  
\begin{tabular}{c}Max time\\algo='g'\end{tabular}& {\tt{ErrMoy}} & {\tt{Errmax}}\\
\hline
40  &   0.30   & 0.004  & 0.34  & 0.008 &  9.5$\small{\times}10^{-10}$  & $10^{-8}$    \\
\hline
100 &   1.99   & 0.007  & 2.54 & 0.014 &   2.5$\small{\times}10^{-9}$       & 4.1$\small{\times}10^{-8}$\\
\hline
200 &   8.09   & 0.012  & 8.96 & 0.018 & 3.8$\small{\times}10^{-9}$        & 3.6$\small{\times}10^{-8}$ \\
\hline
320 &   22.57  & 0.020  & 34.26 & 0.036 &   6.8$\small{\times}10^{-9}$     &  3.4$\small{\times}10^{-8}$ \\
\hline
400 &   45.21  & 0.021  & 669.76 & 0.039 &   1.1$\small{\times}10^{-8}$   & 1.7$\small{\times}10^{-7}$ \\ 
\hline
600 &   128.52 & 0.033  & 2 772.5 & 0.074 &   1.4$\small{\times}10^{-8}$  &1.2$\small{\times}10^{-7}$ \\
\hline
800 &  369.80  & 0.043  & 9 661.8 & 0.076 &  1.7$\small{\times}10^{-8}$   & 9.7$\small{\times}10^{-8}$\\
\hline
\end{tabular}
\caption{Mean and maximal time (in seconds) required to compute the areas of 50 polyhedra
with $4n$ vertices 
for algo='g' and algo='t1'.
The last two columns report respectively the mean and maximal errors.}\label{tablerandompoly}
\end{table}    

We observe that errors are negligible which shows that both algorithms compute
the same areas. Moreover, on all instances
algorithm from \cite{guiguesarxivcg2015} computes all areas
extremely quickly and much quicker than
the algorithm of Section \ref{algotriang1}.
For this latter algorithm,  both the mean
and maximal time required to compute the intersection areas significantly increase
with the number of vertices of the polygone.\\

\par {\textbf{Acknowledgments}} The author's research was 
partially supported by an FGV grant, CNPq grant 307287/2013-0, and 
FAPERJ grants E-26/110.313/2014 and E-26/201.599/2014.

\addcontentsline{toc}{section}{References}
\bibliographystyle{plain}
\bibliography{CDF}

\end{document}

%% file: Triangle.pstex_t
\begin{picture}(0,0)%
\includegraphics{Triangle.pstex}%
\end{picture}%
\setlength{\unitlength}{1450sp}%
\begingroup\makeatletter\ifx\SetFigFont\undefined%
\gdef\SetFigFont#1#2#3#4#5{%
  \reset@font\fontsize{#1}{#2pt}%
  \fontfamily{#3}\fontseries{#4}\fontshape{#5}%
  \selectfont}%
\fi\endgroup%
\begin{picture}(7783,5058)(362,-4615)
\put(1869,-2095){\makebox(0,0)[lb]{\smash{{\SetFigFont{6}{7.2}{\rmdefault}{\mddefault}{\updefault}{\color[rgb]{0,0,0}$S_3$}%
}}}}
\put(946,-3481){\makebox(0,0)[lb]{\smash{{\SetFigFont{6}{7.2}{\rmdefault}{\mddefault}{\updefault}{\color[rgb]{0,0,0}$S_1$}%
}}}}
\put(3241,-4426){\makebox(0,0)[lb]{\smash{{\SetFigFont{6}{7.2}{\rmdefault}{\mddefault}{\updefault}{\color[rgb]{0,0,0}$P$}%
}}}}
\put(518,-4381){\makebox(0,0)[lb]{\smash{{\SetFigFont{6}{7.2}{\rmdefault}{\mddefault}{\updefault}{\color[rgb]{0,0,0}$O$}%
}}}}
\put(5446,-3468){\makebox(0,0)[lb]{\smash{{\SetFigFont{6}{7.2}{\rmdefault}{\mddefault}{\updefault}{\color[rgb]{0,0,0}$S_2$}%
}}}}
\end{picture}%

%% file: Triangle_5_0B.pstex_t
\begin{picture}(0,0)%
\includegraphics{Triangle_5_0B.pstex}%
\end{picture}%
\setlength{\unitlength}{1658sp}%
\begingroup\makeatletter\ifx\SetFigFont\undefined%
\gdef\SetFigFont#1#2#3#4#5{%
  \reset@font\fontsize{#1}{#2pt}%
  \fontfamily{#3}\fontseries{#4}\fontshape{#5}%
  \selectfont}%
\fi\endgroup%
\begin{picture}(5878,4584)(2217,-8033)
\put(3121,-4336){\makebox(0,0)[lb]{\smash{{\SetFigFont{7}{8.4}{\rmdefault}{\mddefault}{\updefault}{\color[rgb]{0,0,0}distributed in the triangle}%
}}}}
\put(3121,-4711){\makebox(0,0)[lb]{\smash{{\SetFigFont{7}{8.4}{\rmdefault}{\mddefault}{\updefault}{\color[rgb]{0,0,0}of the left}%
}}}}
\put(4951,-4711){\makebox(0,0)[lb]{\smash{{\SetFigFont{7}{8.4}{\rmdefault}{\mddefault}{\updefault}{\color[rgb]{0,0,0}figure, $P=(5;0)$}%
}}}}
\put(3121,-3961){\makebox(0,0)[lb]{\smash{{\SetFigFont{7}{8.4}{\rmdefault}{\mddefault}{\updefault}{\color[rgb]{0,0,0}Density of $D$ when $X$ is uniformly}%
}}}}
\put(2851,-4486){\rotatebox{90.0}{\makebox(0,0)[lb]{\smash{{\SetFigFont{7}{8.4}{\rmdefault}{\mddefault}{\updefault}{\color[rgb]{0,0,0}$f_D(d)$}%
}}}}}
\end{picture}%

%% file: TriangleIn.pstex_t
\begin{picture}(0,0)%
\includegraphics{TriangleIn.pstex}%
\end{picture}%
\setlength{\unitlength}{1450sp}%
\begingroup\makeatletter\ifx\SetFigFont\undefined%
\gdef\SetFigFont#1#2#3#4#5{%
  \reset@font\fontsize{#1}{#2pt}%
  \fontfamily{#3}\fontseries{#4}\fontshape{#5}%
  \selectfont}%
\fi\endgroup%
\begin{picture}(7783,5058)(362,-4615)
\put(1869,-2095){\makebox(0,0)[lb]{\smash{{\SetFigFont{6}{7.2}{\rmdefault}{\mddefault}{\updefault}{\color[rgb]{0,0,0}$S_3$}%
}}}}
\put(946,-3481){\makebox(0,0)[lb]{\smash{{\SetFigFont{6}{7.2}{\rmdefault}{\mddefault}{\updefault}{\color[rgb]{0,0,0}$S_1$}%
}}}}
\put(518,-4381){\makebox(0,0)[lb]{\smash{{\SetFigFont{6}{7.2}{\rmdefault}{\mddefault}{\updefault}{\color[rgb]{0,0,0}$O$}%
}}}}
\put(5446,-3468){\makebox(0,0)[lb]{\smash{{\SetFigFont{6}{7.2}{\rmdefault}{\mddefault}{\updefault}{\color[rgb]{0,0,0}$S_2$}%
}}}}
\put(2836,-3481){\makebox(0,0)[lb]{\smash{{\SetFigFont{6}{7.2}{\rmdefault}{\mddefault}{\updefault}{\color[rgb]{0,0,0}$P$}%
}}}}
\end{picture}%

%% file: Triangle_4_2B.pstex_t
\begin{picture}(0,0)%
\includegraphics{Triangle_4_2B.pstex}%
\end{picture}%
\setlength{\unitlength}{1658sp}%
\begingroup\makeatletter\ifx\SetFigFont\undefined%
\gdef\SetFigFont#1#2#3#4#5{%
  \reset@font\fontsize{#1}{#2pt}%
  \fontfamily{#3}\fontseries{#4}\fontshape{#5}%
  \selectfont}%
\fi\endgroup%
\begin{picture}(5887,4584)(2123,-8033)
\put(2851,-4486){\rotatebox{90.0}{\makebox(0,0)[lb]{\smash{{\SetFigFont{7}{8.4}{\rmdefault}{\mddefault}{\updefault}{\color[rgb]{0,0,0}$f_D(d)$}%
}}}}}
\put(3826,-3811){\makebox(0,0)[lb]{\smash{{\SetFigFont{7}{8.4}{\rmdefault}{\mddefault}{\updefault}{\color[rgb]{0,0,0}Density of $D$ when $X$ is}%
}}}}
\put(3865,-4186){\makebox(0,0)[lb]{\smash{{\SetFigFont{7}{8.4}{\rmdefault}{\mddefault}{\updefault}{\color[rgb]{0,0,0}uniformly distributed in the}%
}}}}
\put(4201,-4561){\makebox(0,0)[lb]{\smash{{\SetFigFont{7}{8.4}{\rmdefault}{\mddefault}{\updefault}{\color[rgb]{0,0,0}triangle of the left figure}%
}}}}
\put(4201,-4936){\makebox(0,0)[lb]{\smash{{\SetFigFont{7}{8.4}{\rmdefault}{\mddefault}{\updefault}{\color[rgb]{0,0,0}$P=(4;2)$}%
}}}}
\end{picture}%

%% file: Rectangle.pstex_t
\begin{picture}(0,0)%
\includegraphics{Rectangle.pstex}%
\end{picture}%
\setlength{\unitlength}{1450sp}%
\begingroup\makeatletter\ifx\SetFigFont\undefined%
\gdef\SetFigFont#1#2#3#4#5{%
  \reset@font\fontsize{#1}{#2pt}%
  \fontfamily{#3}\fontseries{#4}\fontshape{#5}%
  \selectfont}%
\fi\endgroup%
\begin{picture}(7758,5058)(397,-4615)
\put(1846,-3076){\makebox(0,0)[lb]{\smash{{\SetFigFont{6}{7.2}{\rmdefault}{\mddefault}{\updefault}{\color[rgb]{0,0,0}$S_1$}%
}}}}
\put(6346,-3076){\makebox(0,0)[lb]{\smash{{\SetFigFont{6}{7.2}{\rmdefault}{\mddefault}{\updefault}{\color[rgb]{0,0,0}$S_2$}%
}}}}
\put(6346,-781){\makebox(0,0)[lb]{\smash{{\SetFigFont{6}{7.2}{\rmdefault}{\mddefault}{\updefault}{\color[rgb]{0,0,0}$S_3$}%
}}}}
\put(1846,-781){\makebox(0,0)[lb]{\smash{{\SetFigFont{6}{7.2}{\rmdefault}{\mddefault}{\updefault}{\color[rgb]{0,0,0}$S_4$}%
}}}}
\put(1418,-3976){\makebox(0,0)[lb]{\smash{{\SetFigFont{6}{7.2}{\rmdefault}{\mddefault}{\updefault}{\color[rgb]{0,0,0}$P$}%
}}}}
\put(518,-4381){\makebox(0,0)[lb]{\smash{{\SetFigFont{6}{7.2}{\rmdefault}{\mddefault}{\updefault}{\color[rgb]{0,0,0}$O$}%
}}}}
\end{picture}%

%% file: Rectangle_1_1B.pstex_t
\begin{picture}(0,0)%
\includegraphics{Rectangle_1_1B.pstex}%
\end{picture}%
\setlength{\unitlength}{1658sp}%
\begingroup\makeatletter\ifx\SetFigFont\undefined%
\gdef\SetFigFont#1#2#3#4#5{%
  \reset@font\fontsize{#1}{#2pt}%
  \fontfamily{#3}\fontseries{#4}\fontshape{#5}%
  \selectfont}%
\fi\endgroup%
\begin{picture}(5950,4584)(2123,-8033)
\put(2626,-3886){\makebox(0,0)[lb]{\smash{{\SetFigFont{7}{8.4}{\rmdefault}{\mddefault}{\updefault}{\color[rgb]{0,0,0}$f_D(d)$}%
}}}}
\put(3451,-6661){\makebox(0,0)[lb]{\smash{{\SetFigFont{7}{8.4}{\rmdefault}{\mddefault}{\updefault}{\color[rgb]{0,0,0}Density of $D$ when $X$ is}%
}}}}
\put(6451,-3886){\makebox(0,0)[lb]{\smash{{\SetFigFont{7}{8.4}{\rmdefault}{\mddefault}{\updefault}{\color[rgb]{0,0,0}$P=(1;1)$}%
}}}}
\put(3451,-7111){\makebox(0,0)[lb]{\smash{{\SetFigFont{7}{8.4}{\rmdefault}{\mddefault}{\updefault}{\color[rgb]{0,0,0}uniformly distributed}%
}}}}
\put(3451,-7561){\makebox(0,0)[lb]{\smash{{\SetFigFont{7}{8.4}{\rmdefault}{\mddefault}{\updefault}{\color[rgb]{0,0,0}rectangle of the left figure}%
}}}}
\put(7105,-7111){\makebox(0,0)[lb]{\smash{{\SetFigFont{7}{8.4}{\rmdefault}{\mddefault}{\updefault}{\color[rgb]{0,0,0}in the }%
}}}}
\end{picture}%

%% file: RectangleIn.pstex_t
\begin{picture}(0,0)%
\includegraphics{RectangleIn.pstex}%
\end{picture}%
\setlength{\unitlength}{1450sp}%
\begingroup\makeatletter\ifx\SetFigFont\undefined%
\gdef\SetFigFont#1#2#3#4#5{%
  \reset@font\fontsize{#1}{#2pt}%
  \fontfamily{#3}\fontseries{#4}\fontshape{#5}%
  \selectfont}%
\fi\endgroup%
\begin{picture}(7758,5058)(397,-4615)
\put(1846,-3076){\makebox(0,0)[lb]{\smash{{\SetFigFont{6}{7.2}{\rmdefault}{\mddefault}{\updefault}{\color[rgb]{0,0,0}$S_1$}%
}}}}
\put(6346,-3076){\makebox(0,0)[lb]{\smash{{\SetFigFont{6}{7.2}{\rmdefault}{\mddefault}{\updefault}{\color[rgb]{0,0,0}$S_2$}%
}}}}
\put(6346,-781){\makebox(0,0)[lb]{\smash{{\SetFigFont{6}{7.2}{\rmdefault}{\mddefault}{\updefault}{\color[rgb]{0,0,0}$S_3$}%
}}}}
\put(1846,-781){\makebox(0,0)[lb]{\smash{{\SetFigFont{6}{7.2}{\rmdefault}{\mddefault}{\updefault}{\color[rgb]{0,0,0}$S_4$}%
}}}}
\put(518,-4381){\makebox(0,0)[lb]{\smash{{\SetFigFont{6}{7.2}{\rmdefault}{\mddefault}{\updefault}{\color[rgb]{0,0,0}$O$}%
}}}}
\put(3736,-2131){\makebox(0,0)[lb]{\smash{{\SetFigFont{6}{7.2}{\rmdefault}{\mddefault}{\updefault}{\color[rgb]{0,0,0}$P$}%
}}}}
\end{picture}%

%% file: Rectangle_6_5B.pstex_t
\begin{picture}(0,0)%
\includegraphics{Rectangle_6_5B.pstex}%
\end{picture}%
\setlength{\unitlength}{1658sp}%
\begingroup\makeatletter\ifx\SetFigFont\undefined%
\gdef\SetFigFont#1#2#3#4#5{%
  \reset@font\fontsize{#1}{#2pt}%
  \fontfamily{#3}\fontseries{#4}\fontshape{#5}%
  \selectfont}%
\fi\endgroup%
\begin{picture}(5887,4584)(2123,-8033)
\put(2551,-3886){\makebox(0,0)[lb]{\smash{{\SetFigFont{7}{8.4}{\rmdefault}{\mddefault}{\updefault}{\color[rgb]{0,0,0}$f_D(d)$}%
}}}}
\put(6301,-3886){\makebox(0,0)[lb]{\smash{{\SetFigFont{7}{8.4}{\rmdefault}{\mddefault}{\updefault}{\color[rgb]{0,0,0}$P=(6;5)$}%
}}}}
\put(3001,-6886){\makebox(0,0)[lb]{\smash{{\SetFigFont{7}{8.4}{\rmdefault}{\mddefault}{\updefault}{\color[rgb]{0,0,0}Density of $D$ when $X$ is}%
}}}}
\put(3001,-7261){\makebox(0,0)[lb]{\smash{{\SetFigFont{7}{8.4}{\rmdefault}{\mddefault}{\updefault}{\color[rgb]{0,0,0}uniformly distributed in the}%
}}}}
\put(3001,-7636){\makebox(0,0)[lb]{\smash{{\SetFigFont{7}{8.4}{\rmdefault}{\mddefault}{\updefault}{\color[rgb]{0,0,0}rectangle of the left figure}%
}}}}
\end{picture}%

%% file: Poly_Ext.pstex_t
\begin{picture}(0,0)%
\includegraphics{Poly_Ext.pstex}%
\end{picture}%
\setlength{\unitlength}{1450sp}%
\begingroup\makeatletter\ifx\SetFigFont\undefined%
\gdef\SetFigFont#1#2#3#4#5{%
  \reset@font\fontsize{#1}{#2pt}%
  \fontfamily{#3}\fontseries{#4}\fontshape{#5}%
  \selectfont}%
\fi\endgroup%
\begin{picture}(7758,5058)(397,-4615)
\put(518,-4381){\makebox(0,0)[lb]{\smash{{\SetFigFont{6}{7.2}{\rmdefault}{\mddefault}{\updefault}{\color[rgb]{0,0,0}$O$}%
}}}}
\put(901,-3571){\makebox(0,0)[lb]{\smash{{\SetFigFont{6}{7.2}{\rmdefault}{\mddefault}{\updefault}{\color[rgb]{0,0,0}$S_1$}%
}}}}
\put(1846,-3931){\makebox(0,0)[lb]{\smash{{\SetFigFont{6}{7.2}{\rmdefault}{\mddefault}{\updefault}{\color[rgb]{0,0,0}$S_2$}%
}}}}
\put(3106,-3571){\makebox(0,0)[lb]{\smash{{\SetFigFont{6}{7.2}{\rmdefault}{\mddefault}{\updefault}{\color[rgb]{0,0,0}$S_3$}%
}}}}
\put(4051,-3976){\makebox(0,0)[lb]{\smash{{\SetFigFont{6}{7.2}{\rmdefault}{\mddefault}{\updefault}{\color[rgb]{0,0,0}$S_4$}%
}}}}
\put(4546,-3031){\makebox(0,0)[lb]{\smash{{\SetFigFont{6}{7.2}{\rmdefault}{\mddefault}{\updefault}{\color[rgb]{0,0,0}$S_5$}%
}}}}
\put(4141,-1276){\makebox(0,0)[lb]{\smash{{\SetFigFont{6}{7.2}{\rmdefault}{\mddefault}{\updefault}{\color[rgb]{0,0,0}$S_7$}%
}}}}
\put(2296,-1681){\makebox(0,0)[lb]{\smash{{\SetFigFont{6}{7.2}{\rmdefault}{\mddefault}{\updefault}{\color[rgb]{0,0,0}$S_8$}%
}}}}
\put(901,-2626){\makebox(0,0)[lb]{\smash{{\SetFigFont{6}{7.2}{\rmdefault}{\mddefault}{\updefault}{\color[rgb]{0,0,0}$S_9$}%
}}}}
\put(1756,-3076){\makebox(0,0)[lb]{\smash{{\SetFigFont{6}{7.2}{\rmdefault}{\mddefault}{\updefault}{\color[rgb]{0,0,0}$S_{10}$}%
}}}}
\put(2791,-4426){\makebox(0,0)[lb]{\smash{{\SetFigFont{6}{7.2}{\rmdefault}{\mddefault}{\updefault}{\color[rgb]{0,0,0}$P$}%
}}}}
\put(3691,-2626){\makebox(0,0)[lb]{\smash{{\SetFigFont{6}{7.2}{\rmdefault}{\mddefault}{\updefault}{\color[rgb]{0,0,0}$S_6$}%
}}}}
\end{picture}%

%% file: Poly_40B.pstex_t
\begin{picture}(0,0)%
\includegraphics{Poly_40B.pstex}%
\end{picture}%
\setlength{\unitlength}{1658sp}%
\begingroup\makeatletter\ifx\SetFigFont\undefined%
\gdef\SetFigFont#1#2#3#4#5{%
  \reset@font\fontsize{#1}{#2pt}%
  \fontfamily{#3}\fontseries{#4}\fontshape{#5}%
  \selectfont}%
\fi\endgroup%
\begin{picture}(5887,4584)(2123,-8033)
\put(2626,-3886){\makebox(0,0)[lb]{\smash{{\SetFigFont{7}{8.4}{\rmdefault}{\mddefault}{\updefault}{\color[rgb]{0,0,0}$f_D(d)$}%
}}}}
\put(3151,-7186){\makebox(0,0)[lb]{\smash{{\SetFigFont{7}{8.4}{\rmdefault}{\mddefault}{\updefault}{\color[rgb]{0,0,0}uniformly distributed in}%
}}}}
\put(3151,-7561){\makebox(0,0)[lb]{\smash{{\SetFigFont{7}{8.4}{\rmdefault}{\mddefault}{\updefault}{\color[rgb]{0,0,0}polyhedron of the left figure}%
}}}}
\put(3265,-6811){\makebox(0,0)[lb]{\smash{{\SetFigFont{7}{8.4}{\rmdefault}{\mddefault}{\updefault}{\color[rgb]{0,0,0}Density of $D$ when $X$}%
}}}}
\put(7051,-7186){\makebox(0,0)[lb]{\smash{{\SetFigFont{7}{8.4}{\rmdefault}{\mddefault}{\updefault}{\color[rgb]{0,0,0}the}%
}}}}
\put(6751,-6811){\makebox(0,0)[lb]{\smash{{\SetFigFont{7}{8.4}{\rmdefault}{\mddefault}{\updefault}{\color[rgb]{0,0,0}is}%
}}}}
\put(6433,-3886){\makebox(0,0)[lb]{\smash{{\SetFigFont{7}{8.4}{\rmdefault}{\mddefault}{\updefault}{\color[rgb]{0,0,0}$P=(4;0)$}%
}}}}
\end{picture}%

%% file: Poly_Int.pstex_t
\begin{picture}(0,0)%
\includegraphics{Poly_Int.pstex}%
\end{picture}%
\setlength{\unitlength}{1450sp}%
\begingroup\makeatletter\ifx\SetFigFont\undefined%
\gdef\SetFigFont#1#2#3#4#5{%
  \reset@font\fontsize{#1}{#2pt}%
  \fontfamily{#3}\fontseries{#4}\fontshape{#5}%
  \selectfont}%
\fi\endgroup%
\begin{picture}(7758,5058)(397,-4615)
\put(518,-4381){\makebox(0,0)[lb]{\smash{{\SetFigFont{6}{7.2}{\rmdefault}{\mddefault}{\updefault}{\color[rgb]{0,0,0}$O$}%
}}}}
\put(901,-3571){\makebox(0,0)[lb]{\smash{{\SetFigFont{6}{7.2}{\rmdefault}{\mddefault}{\updefault}{\color[rgb]{0,0,0}$S_1$}%
}}}}
\put(1846,-3931){\makebox(0,0)[lb]{\smash{{\SetFigFont{6}{7.2}{\rmdefault}{\mddefault}{\updefault}{\color[rgb]{0,0,0}$S_2$}%
}}}}
\put(4051,-3976){\makebox(0,0)[lb]{\smash{{\SetFigFont{6}{7.2}{\rmdefault}{\mddefault}{\updefault}{\color[rgb]{0,0,0}$S_4$}%
}}}}
\put(4546,-3031){\makebox(0,0)[lb]{\smash{{\SetFigFont{6}{7.2}{\rmdefault}{\mddefault}{\updefault}{\color[rgb]{0,0,0}$S_5$}%
}}}}
\put(3691,-2626){\makebox(0,0)[lb]{\smash{{\SetFigFont{6}{7.2}{\rmdefault}{\mddefault}{\updefault}{\color[rgb]{0,0,0}$S_6$}%
}}}}
\put(4096,-1276){\makebox(0,0)[lb]{\smash{{\SetFigFont{6}{7.2}{\rmdefault}{\mddefault}{\updefault}{\color[rgb]{0,0,0}$S_7$}%
}}}}
\put(2296,-1681){\makebox(0,0)[lb]{\smash{{\SetFigFont{6}{7.2}{\rmdefault}{\mddefault}{\updefault}{\color[rgb]{0,0,0}$S_8$}%
}}}}
\put(901,-2626){\makebox(0,0)[lb]{\smash{{\SetFigFont{6}{7.2}{\rmdefault}{\mddefault}{\updefault}{\color[rgb]{0,0,0}$S_9$}%
}}}}
\put(1756,-3076){\makebox(0,0)[lb]{\smash{{\SetFigFont{6}{7.2}{\rmdefault}{\mddefault}{\updefault}{\color[rgb]{0,0,0}$S_{10}$}%
}}}}
\put(2746,-2581){\makebox(0,0)[lb]{\smash{{\SetFigFont{6}{7.2}{\rmdefault}{\mddefault}{\updefault}{\color[rgb]{0,0,0}$P$}%
}}}}
\put(3106,-3571){\makebox(0,0)[lb]{\smash{{\SetFigFont{6}{7.2}{\rmdefault}{\mddefault}{\updefault}{\color[rgb]{0,0,0}$S_3$}%
}}}}
\end{picture}%

%% file: Poly_43B.pstex_t
\begin{picture}(0,0)%
\includegraphics{Poly_43B.pstex}%
\end{picture}%
\setlength{\unitlength}{1658sp}%
\begingroup\makeatletter\ifx\SetFigFont\undefined%
\gdef\SetFigFont#1#2#3#4#5{%
  \reset@font\fontsize{#1}{#2pt}%
  \fontfamily{#3}\fontseries{#4}\fontshape{#5}%
  \selectfont}%
\fi\endgroup%
\begin{picture}(5972,4584)(2123,-8033)
\put(2626,-3886){\makebox(0,0)[lb]{\smash{{\SetFigFont{7}{8.4}{\rmdefault}{\mddefault}{\updefault}{\color[rgb]{0,0,0}$f_D(d)$}%
}}}}
\put(6376,-3886){\makebox(0,0)[lb]{\smash{{\SetFigFont{7}{8.4}{\rmdefault}{\mddefault}{\updefault}{\color[rgb]{0,0,0}$P=(4,3)$}%
}}}}
\put(3121,-6736){\makebox(0,0)[lb]{\smash{{\SetFigFont{7}{8.4}{\rmdefault}{\mddefault}{\updefault}{\color[rgb]{0,0,0}Density of $D$ when $X$ is}%
}}}}
\put(2926,-7111){\makebox(0,0)[lb]{\smash{{\SetFigFont{7}{8.4}{\rmdefault}{\mddefault}{\updefault}{\color[rgb]{0,0,0}uniformly distributed in the}%
}}}}
\put(2926,-7486){\makebox(0,0)[lb]{\smash{{\SetFigFont{7}{8.4}{\rmdefault}{\mddefault}{\updefault}{\color[rgb]{0,0,0}polyhedron of the left figure}%
}}}}
\end{picture}%